\documentclass[twocolumn]{aastex61}

\received{May 4, 2019}
\accepted{September 3, 2019}
\submitjournal{ApJ}

\shorttitle{Chemistry in the center of M83}
\shortauthors{Harada et al.}

\begin{document}

\title{Chemical evolution along the circumnuclear ring of M83}

\correspondingauthor{Nanase Harada}
\email{harada@asiaa.sinica.edu.tw}

\author[0000-0002-6824-6627]{Nanase Harada}
\affiliation{Institute of Astronomy and Astrophysics, Academia Sinica, 11F of AS/NTU
Astronomy-Mathematics Building, No.1, Sec. 4, Roosevelt Rd, Taipei 10617, Taiwan,
R.O.C.}

\author[0000-0001-5187-2288]{Kazushi Sakamoto}
\affiliation{Institute of Astronomy and Astrophysics, Academia Sinica, 11F of AS/NTU
Astronomy-Mathematics Building, No.1, Sec. 4, Roosevelt Rd, Taipei 10617, Taiwan,
R.O.C.}

\author[0000-0001-9281-2919]{Sergio Mart\'in}
\affiliation{European Southern Observatory, Alonso de C\'ordova 3107, Vitacura, Santiago, Chile}
\affiliation{Joint ALMA Observatory, Alonso de C\'ordova 3107, Vitacura, Santiago, Chile}

\author[0000-0002-9668-3592]{Yoshimasa Watanabe}
\affil{Faculty of Pure and Applied Sciences, University of Tsukuba, 1-1-1, Tennodai, Tsukuba, Ibaraki 305-8577, Japan}
\affil{Tomonaga Center for the History of the Universe, University of Tsukuba, 
1-1-1 Tennodai,Tsukuba, Ibaraki 305-8571, Japan}
\affil{College of Engineering, Nihon University, 1 Nakagawara, Tokusada, Tamuramachi, Koriyama, Fukushima 963-8642, Japan}

\author{Rebeca Aladro}
\affiliation{Max Planck Institute for Radio Astronomy, Auf dem H\"ugel 69, D-53121 Bonn, Germany}

\author{Denise Riquelme}
\affiliation{Max Planck Institute for Radio Astronomy, Auf dem H\"ugel 69, D-53121 Bonn, Germany}


\author{Akihiko Hirota}
\affiliation{Chile Observatory, National Astronomical Observatory of Japan, Santiago, Chile}
\affiliation{Joint ALMA Observatory, Alonso de C\'ordova 3107, Vitacura, Santiago, Chile}



\begin{abstract}
We report an astrochemical study on the evolution of interstellar molecular clouds and consequent star formation in the center of the barred spiral galaxy M83. 
We used the Atacama Large Millimeter/submillimeter Array (ALMA) to image molecular species indicative of shocks (SiO, CH$_3$OH), 
dense cores (N$_2$H$^+$), and photodissociation regions (CN and CCH), as well as a radio recombination line (H41$\alpha$) tracing active star-forming regions. 
M83 has a circumnuclear gas ring that is joined at two areas by gas streams from the leading-edge gas lanes on the bar. 
We found elevated abundances of the shock and dense-core tracers in one of the orbit-intersecting areas, and {found peaks of CN and H41$\alpha$ downstream}. 
At the other orbit-intersection area, we found similar enhancement of the shock tracers,
 but less variation of other tracers, and no sign of active star formation in the stream. 
We propose that the observed chemical variation or lack of it is due to the presence or absence of collision-induced evolution of molecular clouds and induced star formation.
This work presents the most clear case of the chemical evolution in the circumnuclear rings of barred galaxies, thanks to the ALMA resolution and sensitivity.
\end{abstract}




\section{Introduction} \label{sec:intro}
\subsection{Astrochemistry and properties of the interstellar medium}
The chemical composition of a molecular cloud is a sensitive probe of its physical conditions.
The abundances of various molecules in Galactic star-forming cores are known to evolve with the stages of star formation \citep{2012A&ARv..20...56C}.
The chemistry of a molecular cloud also reflects its environments 
such as the strong UV radiation \citep[e.g., ][]{1997ARA&A..35..179H} or shocks \citep[e.g., ][]{1979ApJS...41..555H}.

The diagnostic power of astrochemistry can also be applied to a collection of molecular clouds in external galaxies.
For example, galaxies with extreme starburst or nuclear activities are expected to have chemical compositions distinct from that in our Galaxy.
To study the characteristic chemical compositions in such galaxies, spectral scan observations 
have been conducted in various types of galaxies both with single-dish telescopes and interferometers
  \citep{2005ApJ...618..259M,2006ApJS..164..450M,2015A&A...579A.101A,2019PASJ..tmp...40T}.
 While single-dish studies have been useful identifying molecular species whose galaxy-wide abundances change according to the types of galaxies,
 spatially-resolved interferometric observations should better highlight the chemistry in regions with most extreme physical conditions.
Thanks to the high sensitivity and angular resolution of Atacama Large Millimeter/sub-millimeter Array (ALMA),
 such astrochemical studies have become practical for various types of galaxies.
 In particular, relationship between the galactic structures and the chemistry within galactic center regions has been revealed in multi-species astrochemical observations using ALMA
  \citep[e.g., ][]{2014PASJ...66...75T,2015PASJ...67....8N,2015A&A...573A.116M,2015ApJ...801...63M}.

With spatially-resolved astrochemistry in an external galaxy, one can effectively study the time evolution of the chemical and physical properties of the interstellar medium (ISM)
in response to large-scale gas dynamics. This is because a spatial variation of gas properties in a gas flow can be interpreted as a time evolution of the ISM in the flow.

 \subsection{Galactic gas dynamics and ISM evolution}
\begin{figure*}
\centering{
\includegraphics[width=0.5\textwidth,trim = 0 0 0 0]{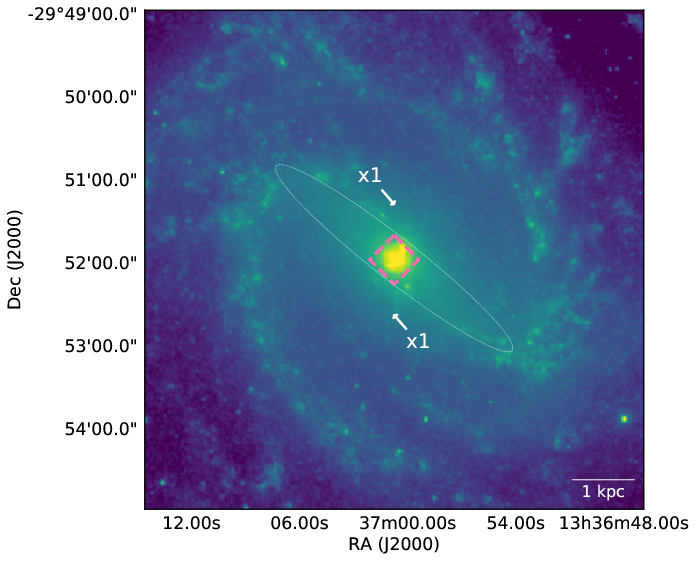}
\includegraphics[width=0.45\textwidth,trim = 0 0 0 0]{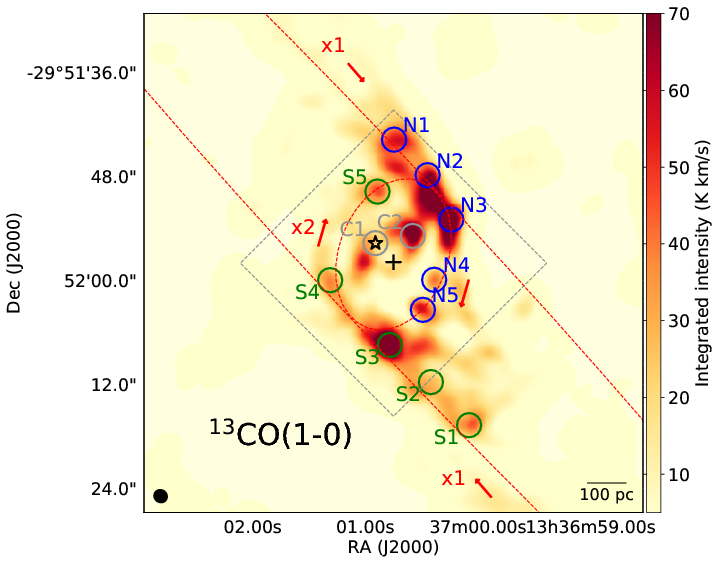}
}
\caption{(Left) Spitzer IRAC 3.6\,$\mu m$ image of M83 from \citet{2009ApJ...703..517D}. The target area of our observations are indicated as a pink dashed rectangule.
An example of $x_1$ (bar) orbits is shown as a white ellipse.
(Right) The velocity-integrated intensity map of $^{13}$CO($1-0$). This is to show the structure of circumnuclear ring and analyzed positions.  The target area of our observations is indicated as a grey dashed rectangle.
Positions analyzed in this paper are indicated as blue (N1-5), green (S1-5), and grey (C1-2) circles.
The star indicates the position of nucleus \citep{2008MNRAS.385.1110H}, and the plus sign shows the location of the center of symmetry.
The synthesized beam is shown as a black ellipse on the lower left corner. 
Examples of $x_1$ and $x_2$ (circumnuclear ring) orbits are shown with red dashed ellipses. 
There is the same map in Appendix \ref{sec:app_mom0} with contours.
The ``typical" value of rms is $\sigma_{\rm N1} = 0.57$ K km s$^{-1}$ in this map.}\label{fig:overview}
\end{figure*}

 A galactic center often hosts a significant fraction of star formation in the galaxy,
 and the gas supply necessary for the star formation is closely related to galactic gas dynamics. 
 In a barred spiral galaxy, the interstellar gas can efficiently flow 
 into the central region because of the shocks on the bar and gravitational torque on the gas in the bar \citep[][]{1976Ap&SS..43..491S,1977Natur.266..607M,1985A&A...150..327C}. 
 Consequently, barred galaxies tend to have higher concentration of molecular gas to their centers \citep{1999ApJ...525..691S},
 and higher star formation activities at the galactic centers \citep{1997ApJ...487..591H}.
 The outline of the relevant gas dynamics is illustrated in Figure~\ref{fig:overview}. The gas in the bar is, outside the central
 region, on the so-called $x_1$ orbits elongated along the bar (Figure~\ref{fig:overview} left).
 The leading-edge gas lanes frequently seen in galactic bars are on these orbits.
 Near the center of the bar, there is often a ring-like structure on the so-called $x_2$ orbits. This structure is sometimes referred to as the (circum-)nuclear 
 ring \citep[][see Figure~\ref{fig:overview} right]{2010MNRAS.402.2462C}.
 Simulations have shown that a circumnuclear ring connected to the leading-edge lanes on the bar naturally result from the non-axisymmetric gravitational potential \citep[][]{1992MNRAS.259..345A}.
  When the gas  flows from the bar through the leading-edge lanes into the circumnuclear ring, collisions between molecular clouds 
  on the $x_1$ and $x_2$ orbits should frequently occur at 
  the accumulation points near orbit intersections. These collisions can unleash star formation events.
    
   Molecular cloud collisions have been proposed as a possible mechanism to trigger star formation \citep{1998ASPC..148..150E}.
  There are many pieces of observational evidence pointing to cloud collisions promoting formation of massive stars in 
  the Galactic spiral-arm regions \citep[e.g., ][]{2009ApJ...696L.115F,2011ApJ...738...46T,2014ApJ...780...36F,2015ApJ...806....7T,2018ApJ...859..166F}
  and the Galactic Center clouds \citep{2015PASJ...67..109T}.
  Magneto-hydrodynamic simulations have also confirmed that collisions can trigger massive star formation \citep{2018PASJ...70S..53I}. 
  While observations of collision-induced star formation in the Galaxy provide detailed views of individual collision sites,
  they provide the information only at the current state of those objects.
In contrast, nearby galaxies, especially face-on galaxies, provide an ideal setting to study time evolution of the ISM through cloud collisions as spatial variation along the gas orbits.

  Collision-induced star formation at the orbit intersection can appear in the form of gradients in stellar ages.
  \citet{2008AJ....135..479B} proposed that two possible modes of star formation in circumuclear rings can cause different distributions of stellar ages.
  One mode is caused by the gravitational collapse of molecular clouds after certain densities are reached in circumnuclear rings
  \citep{1994ApJ...425L..73E}. In this scenario, there is no dependence of star formation on the azimuthal angle on the circumnuclear ring,
  and stellar clusters with a wide range of ages should be randomly located.
 Another scenario is that the star formation is triggered by the collision between the gas on the bar orbit and the gas on the circumnuclear ring.
  In this mode, young star clusters should preferentially exist at the orbit intersections, and there should an stellar age gradient along the stream.
  In the sample of \citet{2008AJ....135..479B}, three out of five galaxies showed such an age gradient.
  In a larger sample, \citet{2008ApJS..174..337M} found that such an age gradient is present in about a half of their sample of 20 barred galaxies
  with circumnuclear rings.
  
  The interaction of the gas on the circumnuclear ring and on the bar is also expected to alter the properties of the ISM.
  \citet{1999ApJ...511..157K} found that the HCN/CO ratio, commonly used as a proxy of the fraction of dense molecular gas, is higher slightly downstream 
  from the orbit intersection in NGC 6951. Even further downstream, there is an H$\alpha$ peak indicating active star formation. 
  Although the higher-resolution follow up study by 
  \citet{2007A&A...468L..63K} showed weaker enhancement of HCN/CO than in \citet{1999ApJ...511..157K}, there still was the general trend of higher HCN/CO downstream
  although \citet{2007A&A...468L..63K} interpreted those enhancements to be random and unrelated to the orbits.
  Such a change in the ISM properties was found also in NGC 7552 with higher HCN/CO observed at the orbit intersections \citep{2013ApJ...768...57P}. 
  On the other hand, the excitation conditions of the molecular gas are not observed to change at the orbit intersection in the case of NGC 1097 \citep{2011ApJ...736..129H}. 
  
\subsection{Previous astrochemical studies in the circumnuclear rings and this work}
 Since the chemical composition is sensitive to various ISM conditions, the astrochemical approach using many species can better probe the ISM evolution
in the circumnuclear rings of galaxies compared to the approach using a small number of line ratios.
 Such an astrochemical study was first conducted by \citet{2005ApJ...618..259M} in the center of the barred spiral galaxy IC 342.
 They observed eight molecules (C$_2$H, C$^{34}$S, N$_2$H$^+$, CH$_3$OH, HNCO, HNC, HC$_3$N, and SO) in IC 342 using the Owens Valley Millimeter Array, 
 and showed that various molecular species peak at different locations.
They proposed that shock tracers, CH$_3$OH and HNCO, are enhanced at locations of shocks at the intersections of the circumnuclear ring and the bar leading-edge gas lanes, that
N$_2$H$^+$ traces the overall dense gas,
and that a photodissociation region (PDR) tracer C$_2$H is enhanced at locations with high star formation rate at the center of the nuclear ring.

 It is worthwhile to follow up the astrochemical variation along the circumnuclear ring with the much improved sensitivity and resolution of ALMA.
 To study the variation of molecular composition caused by the galactic structure and dynamics,
  M83 provides an excellent laboratory due to its proximity \citep[$d \sim 4.5\,$Mpc ($1'' = 22\,$pc); ][]{2003ApJ...590..256T}, accessibility to ALMA,
 its face-on configuration, and presence of a circumnuclear ring. It is a barred spiral galaxy with the morphological type of SAB(s)c \citep{deVaucouleurs1991}.
 The galactic bar has the length of 200$''$ (4.4 kpc) on the sky as seen from the $3.6\,\mu$m image in Figure \ref{fig:overview} (left),
 and the structures of spiral arms are also clearly seen there.
 Such structures have also been seen with CO($J=1-0$) observations \citep{2004A&A...413..505L,2014PASJ...66...46H} to trace molecular gas.
 At the center, there is a circumnuclear ring with the size of 12-16$''$ on the sky (260-350 pc) as shown in Figure \ref{fig:overview} (right).

 In this paper, we report the first spatially resolved, interferometric astrochemical study in the central region of M83 using ALMA observations. 
 Using astrochemical tracers, we study changes in the ISM properties most likely due to collisions of molecular clouds
  at the orbit intersections of the bar leading edge and the circumnuclear ring.
 The observed changes suggest eventual collision-induced star formation.
 The organization of this paper goes as follows. In Section \ref{sec:obs}, we summarize our observations and analysis method.
 Next, our results are summarized in Section \ref{sec:res} to show the chemical variation along the circumnuclear ring.
 We propose an explanation for such chemical variation as well as other possible scenarios in Section \ref{sec:disc}.
 Finally, a summary is given in Section \ref{sec:sum}.

\section{Observations and data reduction} \label{sec:obs}
Our observations covering six tunings were conducted during ALMA cycle 4 using both the 12-m array and the 7-m array (project code: 2016.1.00164.S; PI: Harada).
We targeted the entire circumnuclear ring, and we mosaicked the area of $25\arcsec \times 25\arcsec$ (see Figure \ref{fig:overview} left) for Band 6 and Band 7, whose primary beam sizes are smaller than our targeted area. 
We supplemented our data with ALMA archival data 2012.1.00762.S (PI: Hirota) for CN($N=1-0$) and 2015.1.01177.S (PI: Longmore) for CCH($N=1-0$) and SiO($2-1$) lines.
The former obervations included the 7-m array while the latter only had the 12-m array.
The observational parameters are summarized in Table \ref{tab:obs_param}.
The data cubes from 2012.1.00762.S were created in the data reduction procedure described in \citet{2018PASJ...70...73H},
and smoothed to the velocity resolution of 10.6 km s$^{-1}$. 
The data of 2016.1.00164.S and 2015.1.01177.S were reduced with the following processes.
Measurement sets were calibrated with the CASA pipeline versions 4.6, 4.7 or 4.7.2.
Imaging was conducted with CASA version 5.4 using the task tclean, and the imaging parameters used for individual lines are listed in Table \ref{tab:im_param}.
All the cubes were smoothed to 10 km\,s$^{-1}$ resolution except for CN(1-0) lines smoothed to 10.6 km s$^{-1}$.
Missing flux was assessed for the lines in Band 3 by comparing our line fluxes with those from the IRAM 30-meter line survey in the 3-mm band by \citet{2015A&A...579A.101A}. 
The flux recovery rates are shown in Table \ref{tab:im_param}.
Most lines have the flux recovery rates higher than 70 \% except for CH$_3$CCH  (63 \%) and HNCO (65 \%).
The missing flux is likely insignificant in our analysis because 
we analyze molecular emission at relatively compact peaks.
The maximum recoverable scale is at least 16$''$ in our Band 7 observations,
and this UV coverage should be enough for our target molecules
(See also Table \ref{tab:obs_param} for the UV range and baseline lengths).
The absolute flux calibration is accurate to 5 \% for ALMA Bands 3, 6, and 7, and to 10 \% at IRAM 30m,
and their errors also contribute to the discrepancy between 
their flux measurements.
No single-dish flux was available to evaluate missing flux for the Band 6 and 7.

\section{Results} \label{sec:res}
\subsection{Molecular emission and kinematics}
Figure \ref{fig:overview} (right) shows the overall molecular gas distribution in the central kpc of M83 in the velocity-integrated map of $^{13}$CO($J=1-0$).
Positions used for our molecular abundance analysis in Section \ref{sec:molabun} are shown as circles.
A circumnuclear ring is clearly seen in this molecular emission (the structure connecting positions N2-5 and S3-5).
 The distribution of $^{13}$CO($J=1-0$) is similar to the ones 
 already shown for $^{12}$CO($J=2-1$) \citep{2004ApJ...616L..59S} and $^{12}$CO($J=1-0$) \citep{2018ApJ...854...90E,2018PASJ...70...73H}.
Previous kinematic studies based on large-scale observations showed that the gas is flowing from the bar  into the circumnuclear ring from the northeast and southwest directions
\citep{2004A&A...422..865L,2014PASJ...66...46H}.
Positions following the gas flow from the northeast direction are labeled as N\# where \# is a number starting from upstream (N1-N5; northeastern stream hereafter),
and the gas flow from the southwest direction are shown as S\# (southwestern stream hereafter; S1-S5).

In addition to the positions in the circumnuclear ring, we analyze two nuclear positions. 
The position of the optical nucleus measured by \citet{1991A&A...243..309G} is indicated as C1, and it is about 3$''$ offset from 
the center of symmetry \citep[e.g., ][]{2000A&A...364L..47T,2008MNRAS.385.1110H}.
Another peak of molecular mass concentration is indicated as C2.
The coordinates of the analyzed positions are listed in Table \ref{tab:coords}.

\begin{figure*}
\centering{
\includegraphics[width=0.45\textwidth,trim = 0 0 0 0]{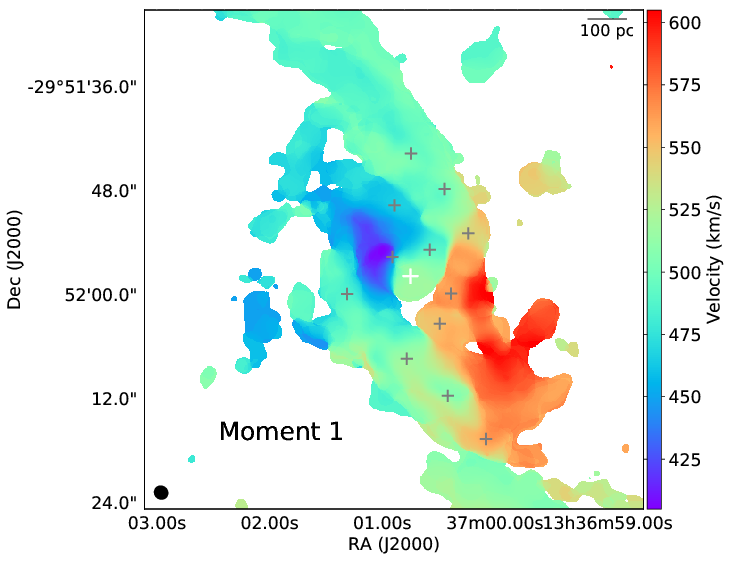}
\includegraphics[width=0.45\textwidth,trim = 0 0 0 0]{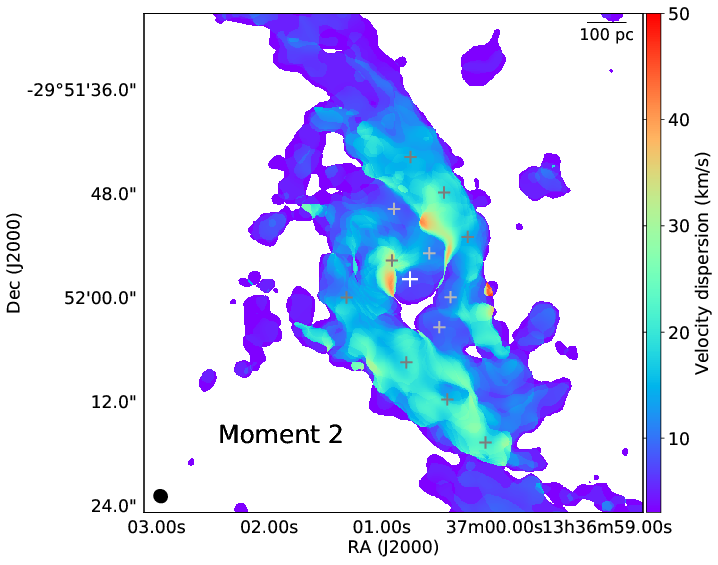}
}
\caption{Moment 1 (mean velocity; left figure) and moment 2 (velocity dispersion; right figure) maps obtained from $^{13}$CO($1-0$) emission. 
The synthesized beam is shown as a black ellipse on the lower left corner. The position of the center of symmetry is indicated as a white plus sign.
Analyzed positions indicated as N1-5, S1-5, and C1-2 in Figure \ref{fig:overview} (right) are shown as grey plus signs.
 \label{fig:mom1mom2}}
\end{figure*}

The rotation of the circumnuclear ring is well-traced in the moment 1 map (Figure \ref{fig:mom1mom2} left).
The velocity dispersion shown in the moment 2 map (Figure \ref{fig:mom1mom2} right) increases from $< 10\,$km s$^{-1}$ in the bar to $\sim 20\,$km s$^{-1}$ when entering 
into the circumnuclear ring at N1 and S1, likely showing the effect of crowding at orbit intersections leading to cloud collisions.
There are two peaks of velocity dispersion with $\sigma \sim 40$ km s$^{-1}$; one is near C1, and the other is a position between C2, N2, and N3. 
The first peak is caused by the motion near the optical nucleus, while the second peak
is likely due to double velocity peaks on the line of sight. 
These double velocity peaks may indicate either two colliding clouds or two clouds that happen to be in the same line of sight,
and it is difficult to distinguish between them.

\subsection{Spatial variation of molecular emission}\label{sec:molemit}
\begin{figure*}
\centering{
\includegraphics[width=0.45\textwidth,trim = 0 0 0 0]{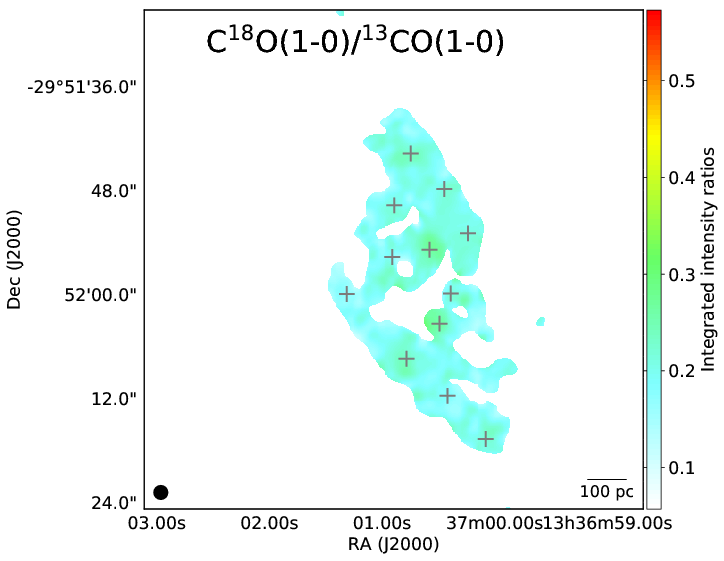}
\includegraphics[width=0.45\textwidth,trim = 0 0 0 0]{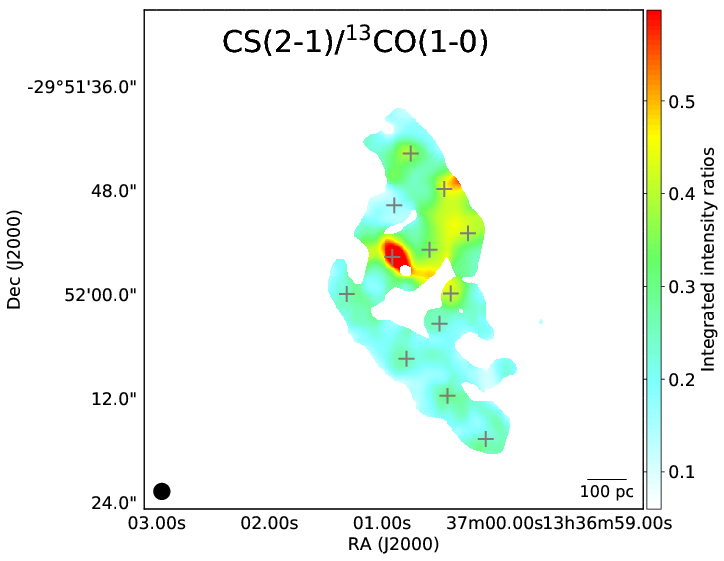}
}
\centering{
\includegraphics[width=0.45\textwidth,trim = 0 0 0 0]{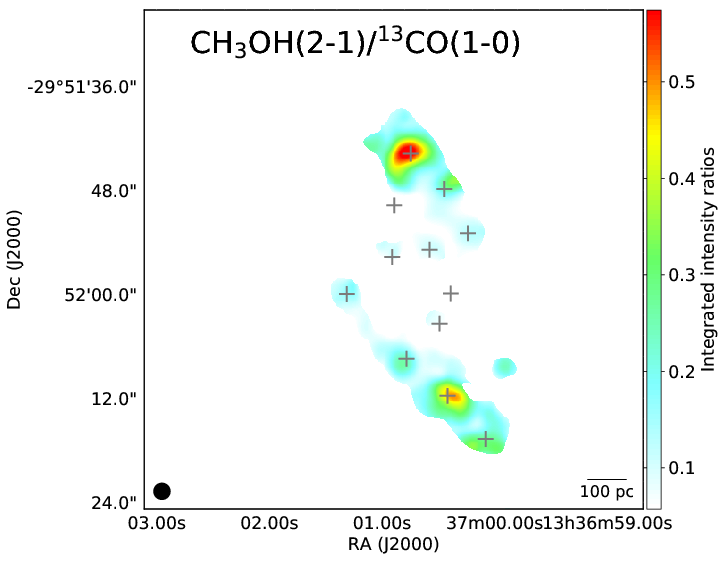}
\includegraphics[width=0.45\textwidth,trim = 0 0 0 0]{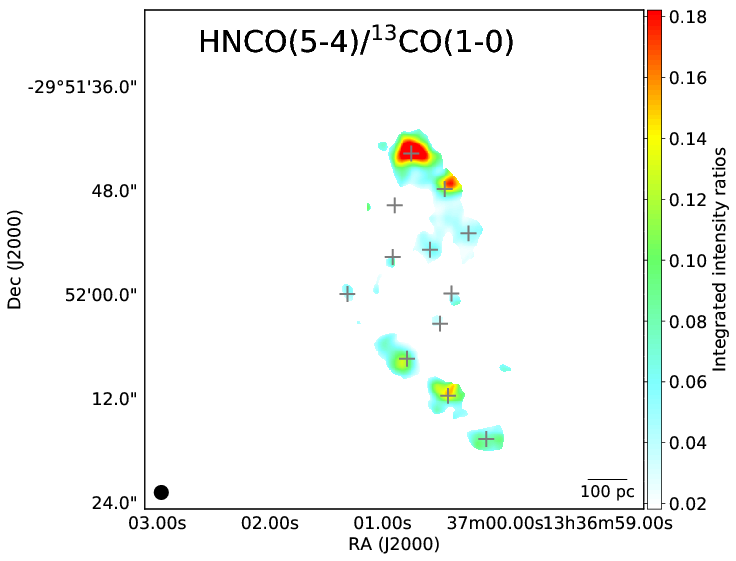}
}
\centering{
\includegraphics[width=0.45\textwidth,trim = 0 0 0 0]{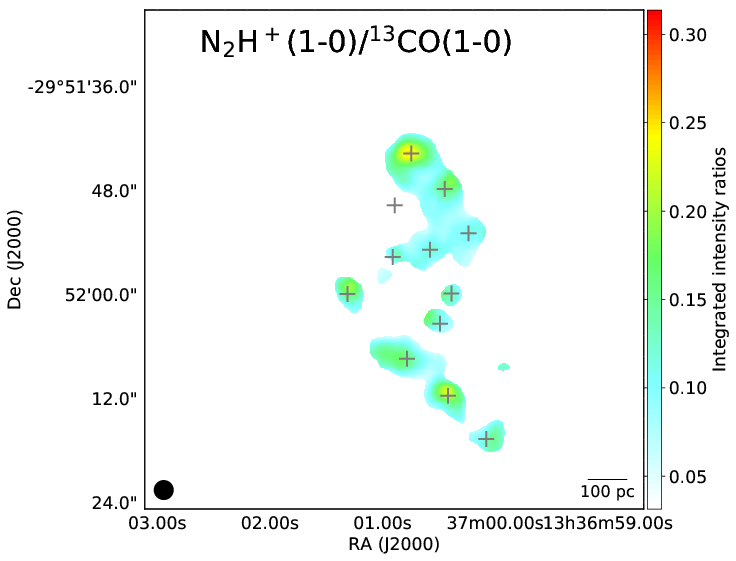}
\includegraphics[width=0.45\textwidth,trim = 0 0 0 0]{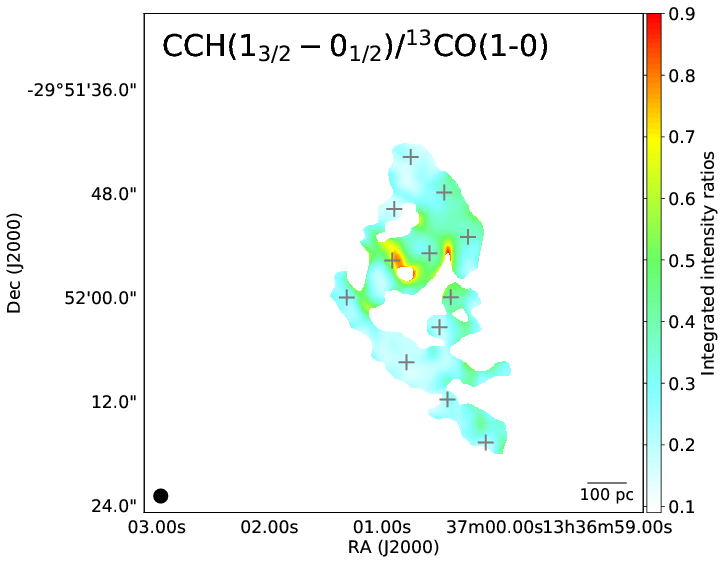}}
\caption{Ratios of line intensities of selected species over that of $^{13}$CO ($1-0$) in Kelvin scale. Beam sizes are convolved to the same resolution, a larger
one between the lines in the numerator and the denominator. All the images are corrected for the primary beam. 
Ratios of the maximum and the minimum value of the color scale are set to be 10. \label{fig:mom0_ratios}}
\end{figure*}

We start from examining the spatial variation of the chemistry with the ratio maps of velocity-integrated intensities in Figure~\ref{fig:mom0_ratios}.
All these ratio maps used $^{13}$CO($1-0$) as the denominator, and were created in the following way.
We first made the velocity-integrated intensity maps (moment 0 maps) shown and described in Appendix \ref{sec:app_mom0}.
The two images for each ratio map were next convolved to the largest beam among the two images.
The image division for ratio was then performed after applying a cutoff at $6\sigma$ and correction for the primary beam to each convolved image.
 
It is evident from Figure \ref{fig:mom0_ratios} that the distribution of emission vary among different molecular species.
Although the C$^{18}$O($1-0$)/$^{13}$CO($1-0$) ratio is almost constant throughout the central region of M83 within a factor of 1.6,
intensity ratios of some other transitions over $^{13}$CO($1-0$) have noticable spatial variation.

On the flow coming from the northeast direction (positions N1-5), the CH$_3$OH ($2_k-1_k$)/$^{13}$CO($1-0$) ratio has a strong peak at position N1,
and the HNCO($5_{0,5}-4_{0,4}$)/$^{13}$CO($1-0$) ratio is enhanced at N1 and N2.
Shocks are able to enhance the gas-phase CH$_3$OH by sputtering off the ice on grains \citep[e.g., ][]{2012MNRAS.421.2786F}.
Enhanced CH$_3$OH emission in the large scale is likely to 
be caused by shocks, because the other mechanism  to enhance CH$_3$OH, thermal evaporation from protostellar
cores, can only act in sub-pc scale \citep[e.g., ][]{2005ApJ...618..259M,2006A&A...450L..13M,2006A&A...455..971R,2018ApJ...855...49H}.
HNCO also has a fast formation route on the grain surface, and its abundance is observationally and theoretically shown to be enhanced in shocked regions
\citep{2000A&A...361.1079Z,2008ApJ...678..245M,2009ApJ...706.1323M,2017A&A...597A..11K}.
This N1 position is where the velocity dispersion increases as the gas flows from the bar into the circumnuclear ring,
and frequent shocks are expected around there.

The N$_2$H$^+$($1-0$)/$^{13}$CO($1-0$) ratio is also enhanced at N1.
Galactic large-scale observations show that N$_2$H$^+$ traces dense cores, even better than conventional ``dense-gas tracers"
 such as HCN and HCO$^+$ do \citep{2017A&A...599A..98P,2017A&A...605L...5K,2017ApJ...845..116W,2019ApJ...871..238H} because HCN and HCO$^+$ can be sub-thermally excited when they are optically thick. Another explanation is that species such as HCN, HCO$^+$, and CS can be excited by
 collisions with electrons \citep{2016ApJ...823..124L,2017ApJ...841...25G}. On the other hand, N$_2$H$^+$ abundance is theoretically known
 to increase in the dense environment without the strong influence of UV photons 
  in the model of a collapsing prestellar core \citep{2001ApJ...552..639A}.
If shocks due to molecular cloud collisions compress gas, this compression increases the dense gas fraction,
and can explain the variation of N$_2$H$^+$($1-0$).

The ratio of CCH($1_{3/2}-0_{1/2}$) over $^{13}$CO($1-0$) does not vary significantly 
along N1-5 although there is a very slight enhancement at N2 and N3.
The CS($2-1$)/$^{13}$CO($1-0$) ratio does not vary significantly either, except for the enhancement at N2. 
Ethynyl radical (CCH) is thought to be a tracer of PDRs \citep{2005A&A...435..885P,2008ApJ...675L..33B,2014A&A...563L...6M}. 
 CS($2-1$) emission should be rather ubiquitous, but CS emission should come more from relatively denser regions than $^{13}$CO($1-0$) emission does,
due to the higher critical density although excitation from collision with electrons can also contribute if the electron fraction is high.

In the gas flow from the southwestern direction, lines that are enhanced at N1 or N2 with respect to $^{13}$CO($1-0$) such as CH$_3$OH ($2_k-1_k$),
 HNCO($5_{0,5}-4_{0,4}$), and N$_2$H$^+$($1-0$), have peaks at S2. 
 Other lines such as CCH($1_{3/2}-0_{1/2}$) and CS($2-1$) do not change significantly with respect to $^{13}$CO($1-0$).
  As for the central positions, the CCH($1_{3/2}-0_{1/2}$)/$^{13}$CO($1-0$) and CS($2-1$)/$^{13}$CO($1-0$) ratios are enhanced around the optical nucleus (C1).
  This enhancement may be caused by the lower optical depth of CCH and CS due to the higher velocity dispersion around C1.
  Because the C1 position is not close to prominent star clusters, enhancement due to PDRs is unlikely.

\subsection{Molecular abundances}\label{sec:molabun}
We derived column densities of species by fitting the spectra at our twelve sampling positions under the assumption of the local thermodynamic equilibrium (LTE).
We used the MADCUBA software (S. Mart\'in et al., 2019, submitted)\footnote{http://cab.inta-csic.es/madcuba/Portada.html}
that accounts for the line optical depths and returns column densities and excitation temperatures for the fitted molecules.
This analysis used data cubes convolved to 2.8$''$ (62 pc), which is the lowest resolution in our data.
While at least two transitions for a species are needed to derive the excitation temperature together with the column density,
 some species satisfied this condition only at some of our sampling positions.
We therefore assumed the following excitation temperatures at positions where only a single transition was detected for each species: 
CS: $8 \pm 2$ K, CCH: $3.5 \pm 1.5$ K, N$_2$H$^+$: $6\pm1$ K,
H$_2$CO: $20 \pm 10$ K, HC$_3$N: $15 \pm 10$ K, CH$_3$CCH: $20 \pm 10$ K.
These temperatures are from positions with successful $T_{ex}$ fitting, and we assumed them to have fixed ranges because excitation temperatures did not vary significantly 
among positions for species with multiple transitions detected at all positions.
For C$^{17}$O, we only observed $J=2-1$ in our settings, and assumed its excitation temperature to be similar to those of C$^{18}$O and $^{13}$CO at each position.
If a species has both ortho and para states such as H$_2$CO, we used the ortho-to-para ratio of 3, which is the value of the statistical weight.
This ortho-to-para ratio is achieved when the thermal equilibrium is achieved and the temperature is high compared with the energy differences 
between ortho and para states.
This assumption of ortho-to-para ratio is justified in the warm environment of our observations, a galactic center in a starburst galaxy.

The upper panel of Figure \ref{fig:col_ratio1} shows the column density ratios of selected species over $^{13}$CO 
for positions in the northeastern stream (N1-5), the southwestern stream (S1-5), and central positions (C1-2).
 The lower panel shows the same values normalized to N3.
For simplicity, we refer to the column density ratios as fractional abundances\footnote{Fractional abundances are defined as ratios of densities of certain species 
per unit volume over molecular hydrogen density.}, assuming that the $^{13}$CO fractional abundance
does not vary among positions.

For the northeastern stream (N1-N5), the change in fractional abundances from N1
can be categorized into monotonic increase, no change, or decrease
along the stream.
Most species decrease in fractional abundances such as CH$_3$OH, N$_2$H$^+$, and HC$_3$N.
The largest change from the upstream to the downstream is seen in CH$_3$OH, and it is by a factor of $\sim 10$
while the change is only a factor of $\sim 2$ for N$_2$H$^+$.
The fractional abundances of C$_2$H, CS, CH$_3$CCH, and H$_2$CO show little change along the northeastern stream
except at N5, where H$_2$CO is about 3 times less abundant than that at other positions.
The fractional abundance of CN is the lowest at N1,  $\sim4$ times higher at N2, and stays constant further downstream.

In contrast to the northeastern stream, most of the species in the southwestern stream (S1-S5) do not show 
simple monotonic increase or decrease in fractional abundances; they rather fluctuate within the stream.
An exception is CH$_3$OH, whose fractional abundance monotonically decreases  along the southwestern stream
as it does in the northeastern stream.

The two peaks in the central region C1 and C2 show little difference from other positions in their chemical compositions.
Exceptions are C1 and C2 having higher CN fractional abundances compared with the peaks in the northeastern and southwestern streams, 
and C2 having the highest CH$_3$CCH fractional abundances among all positions.

\begin{figure*}
\centering{
\includegraphics[width=0.49\textwidth,trim = 0 0 0 0]{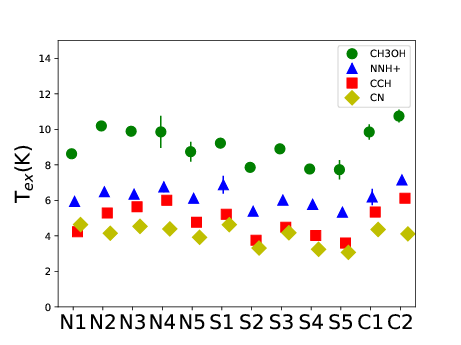}
\includegraphics[width=0.49\textwidth,trim = 0 0 0 0]{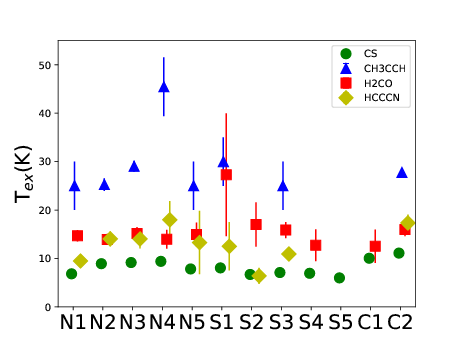}
}
\caption{Excitation temperatures derived from our analysis. Values are not shown if only one transition of the particular molecule was detected. \label{fig:Tex}}
\end{figure*}

Our derived excitation temperatures are shown in Figure \ref{fig:Tex}. 
For most species, the excitation temperatures do not vary significantly from position to position,
and hence intensity ratios are relatively good measures of column density ratios of our observed molecules.
Despite the invariance between positions, there are differences of excitation temperatures between species.
Such variations have also been reported in other external galaxies \citep[e.g., ][]{2006ApJS..164..450M,2011A&A...535A..84A,2018PASJ...70....7N}
This variation between species is likely because each species emit from different components within the observed beam.
Exact values of column densities derived from our results are listed in Tables \ref{tab:colN} - \ref{tab:colC}.

\subsection{Intensity ratios}\label{sec:intratio}
\begin{figure*}
\centering{
\includegraphics[width=0.99\textwidth,trim = 0 0 0 0]{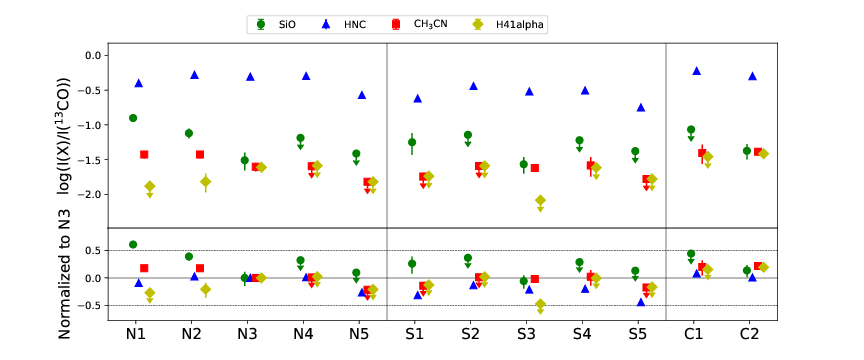}
}
\caption{Intensity ratios of molecular emission over that of $^{13}$CO ($1-0$) (Top panels). Bottom panels show the same 
quantities, but are normalized to the values at N3. \label{fig:intens_ratio1}}
\end{figure*}

We also analyzed species with only one transition in our observations and a radio recombination line H41$\alpha$ for their variation among our twelve positions in 
intensity ratios with respect to $^{13}$CO$(1-0)$.
Figure \ref{fig:intens_ratio1} shows the ratios and ratios normalized to the values at N3, all for the 2.8$''$ resolution. 
HNC does not show evident spatial variation except for lower fractional abundances in the downstream (N5 and S5) by a factor of a few.
The variation of CH$_3$CN fractional abundance is not evident either because there are non-detections in some positions.
Meanwhile, SiO and H41$\alpha$ show hints of spatial variation.
$I({\rm SiO})/I(^{13}{\rm CO})$ is more than a factor of 3 higher at N1 than at N3. 
When comparing S1 and S3, the enhancement of $I({\rm SiO})/I(^{13}{\rm CO})$ at the upper stream, i.e., S1, is less obvious.
SiO is known to be formed when Si is ejected from grains because of fast shocks \citep{1992A&A...254..315M,2008A&A...482..809G,2012MNRAS.421.2786F},
and this variation may be caused by the difference in the frequency of strong shocks.

It is not straightforward to interpret the variation of H41$\alpha$ intensity ratios because H41$\alpha$ is detected only at N2 and N3 and 
we could only get high upper limits for some of other positions.
Yet, the difference between the northeastern stream and southwestern stream can be highlighted by the difference between N3 and S3,
the locations with the highest $^{13}$CO column densities in each stream.
The intensity ratio  $I({\rm H41\alpha})/I(^{13}{\rm CO})$ is more than a factor of 3 higher at N3 than at S3,
with only an upper limit obtained at S3. 
Radio recombination lines from ionized gas trace high-mass star formation. Therefore, this difference suggests a higher star formation rate and a higher star formation 
efficiency in the northeastern stream than in the southwestern stream.
 Observed intensities in the analyzed positions are listed in Table \ref{tab:intens}.

\section{Discussion} \label{sec:disc}
\subsection{A scenario proposed for the chemical variation in the northeastern stream}\label{sec:prop_scenario}
The chemical variation we reported in Sections \ref{sec:molemit} - \ref{sec:intratio} is as follows.
At the upstream positions N1 and N2, CH$_3$OH fractional abundance and the intensity ratio of SiO/$^{13}$CO are higher than at other positions.
These species are known to be enhanced with shocks.
The fractional abundance of N$_2$H$^+$ is also elevated at N1 and N2.
It is known that the N$_2$H$^+$ fractional abundance is enhanced in dense regions.
On the other hand, the CN fractional abundance increases from N1 to N2 by about a factor of 3, and stay at similar values towards downstream,
while CCH fractional abundance does not change along the northeastern stream.
Similar to CCH, CN is known to be a PDR tracer \citep{1993A&A...276..473F}.
The hydrogen recombination line H41$\alpha$ is clearly detected around N2 and N3 although there is no detection at other positions.

Here we note that the changes in derived fractional abundances or intensity may correspond to changes in the fraction of the gas 
that are shocked, dense, or UV-irradiated within the beam.  

\begin{figure*}
\centering{
\includegraphics[width=0.99\textwidth,trim = 0 0 0 0]{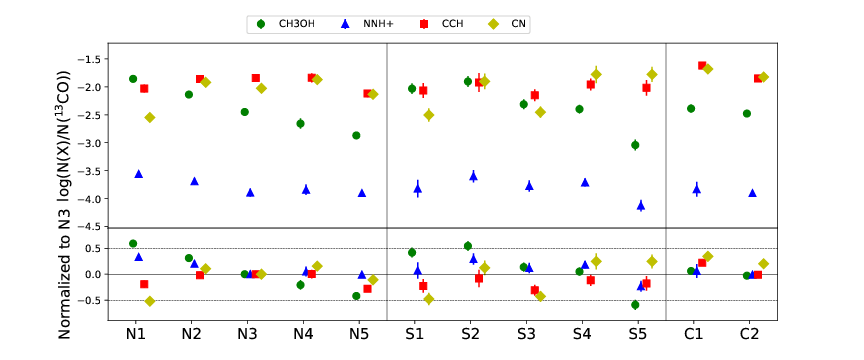}
\includegraphics[width=0.99\textwidth,trim = 0 0 0 0]{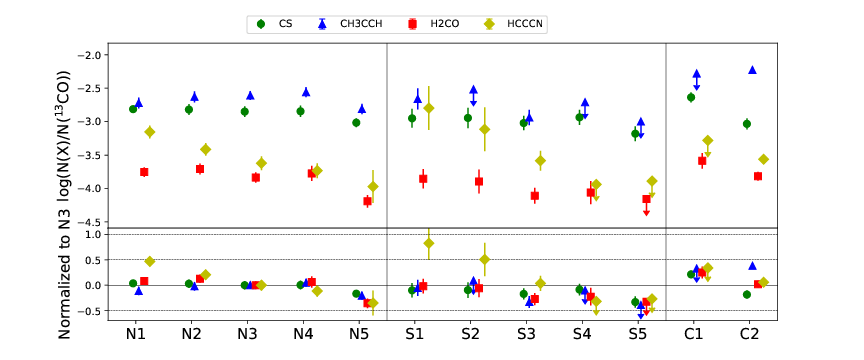}
}
\caption{Column density ratios of molecular species over that of $^{13}$CO for each position (Top panels). Bottom panels show the same 
quantities, but are normalized to the values at N3. \label{fig:col_ratio1}}
\end{figure*}

\begin{figure*}
\centering{
\includegraphics[width=0.38\textwidth,angle=0,trim = 50 50 50 50]{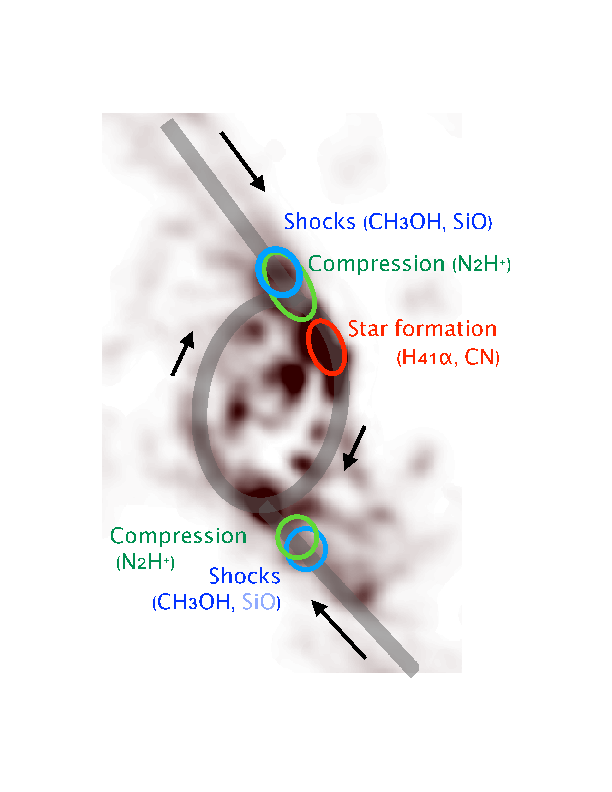}
\includegraphics[width=0.6\textwidth,trim = 0  0 0 0]{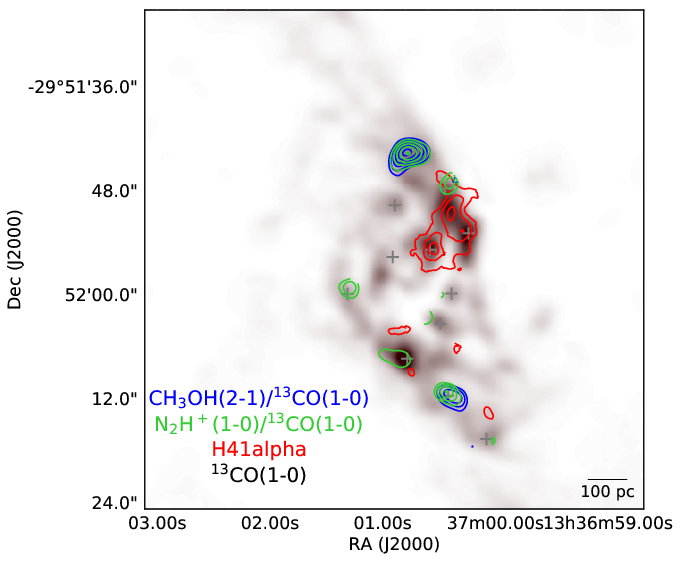}
}
\caption{(Left) A schematic image of our proposed scenario. The background image is a moment 0 map of $^{13}$CO($1-0$). 
(Right) Contours of CH$_3$OH(2-1)/$^{13}$CO(1-0) velocity-integrated intensity ratio shown in blue, N$_2$H$^+$/$^{13}$CO(1-0) velocity-integrated 
intensity ratio in green, and H41$\alpha$ velocity-integrated intensity in red overlaid on a moment 0 map of $^{13}$CO($1-0$). \label{fig:scenario}}
\end{figure*}

From the variation observed in the northeastern stream, we propose the following scenario (Figure~\ref{fig:scenario}).
N1 is the entering point of the gas from the bar orbit into the circumnuclear ring, dominated by frequent hydrodynamical shocks.
These shocks can explain the increase of CH$_3$OH abundance and SiO intensity at N1 and N2.
At these positions, N1 and N2, the velocity dispersion is slightly higher ($\sim 15\,$km s$^{-1}$) than to the north of N1 ($v_{\rm disp} <10\,$km s$^{-1}$).
This is consistent with our inference that there must be shocks at these positions.
These shocks must also have increased the fractional abundance of N$_2$H$^+$ because the compression of 
colliding molecular clouds should increase the gas density. 
This downstream compression may explain the active star formation between N2 and N3 seen in H41$\alpha$ (Figure \ref{fig:col_ratio1}). 
The increase of CN fractional abundance downstream agrees with the higher star formation rate at N2 and N3,
but the invariance of the CCH fractional abundance along the stream cannot be explained if the CCH fractional 
abundance directly reflects the strength of the interstellar radiation field.
There are various claims how the CCH abundance depends on the UV-radiation strength.
\citet{2005ApJ...618..259M}  found in the circumnuclear ring of IC 342 that CCH peaks do not coincide with young ($\sim$ a few Myr old) star-forming regions
in two of the giant moelcular clouds on the ring.
They deduced that the effect of the still embedded young stars are too localized and minimized to have CCH peaks. 
Model calculations by \citet{2013A&A...549A..39A} showed that CCH abundance is rather insensitive to the UV field.
On the other hand, results by \citet{2014A&A...563L...6M} seem to show CCH enhancement over HCN, HCO$^+$, and HNC due to young star formation
in an isolated star-forming galaxy in CIG 638.

\subsection{Timescales of collision-induced star formation}

\subsubsection{Orbital timescale}\label{sec:orbit}
For the proposed scenario of collision-induced star formation to be plausible, the timescale of the gas moving from N1 to N3 should be 
reasonable for star formation to take place. The timescale can be estimated from the orbital timescale and the positional offset between N1 and N3.
We estimated  the orbital timescale for a gas cloud to go around once on the circumnuclear ring as follows.
The difference between the maximum line-of-sight velocity on the circumnuclear ring and the mean line-of-sight velocity of the ring $v_{los}$ is estimated to be $70\pm10$ km s$^{-1}$
on our moment 1 map.
With the inclination angle of 24$^{\circ}$\citep{1981A&AS...44..441C}, the unprojected velocity $v = v_{los}/{\rm sin}(24^{\circ})$ is $172\pm25$ km s$^{-1}$.
From our molecular emission map, the radius of the circumnuclear ring $R$ is estimated to be $250\pm25$ pc.
The angular velocity of the gas in the circumnuclear ring $\Omega$ is $\Omega = v/R = 690 \pm 120$ km s$^{-1}$ kpc$^{-1}$.
Because the pattern speed of the bar $\Omega_{p}$ is $57\pm3$ km s$^{-1}$ kpc$^{-1}$ \citep{2014PASJ...66...46H},
the angular speed of a cloud in the circumnuclear ring with respect to the ring structure 
is $\Omega - \Omega_{p} = 630\pm 120$ km s$^{-1}$ kpc$^{-1}$.
The orbital period $\tau$ is obtained as $\tau = \frac{2\pi}{\Omega - \Omega_{p}} = 10 \pm 2$ Myr.
This orbital period means that it only takes a few Myr for a cloud to travel from the initial shock at N1 to star formation at N3.

\subsubsection{Collision timescale}\label{sec:collision_time}
If star formation is triggered at the orbit intersection through cloud collisions, the collision timescale plus the star formation timescale should be as short as a few Myr for our proposed scenario viable.
The collision timescale of molecular clouds in spiral arm regions is estimated to be 8-10 Myr \citep{2015MNRAS.446.3608D}.
However, we expect the collision timescale to be much shorter in the area of our observations because it is at the galactic center region with much higher number density of molecular clouds,
and because the orbit intersection increases the chance of collision.

An order-of-magnitude estimation of the timescale of collisions can be made by considering the mean free path and the velocity (or velocity dispersion)
of molecular clouds. The mean free path of a cloud is $l= \frac{1}{n_{cl}\sigma}$ where $l$ is the mean free path, $n_{cl}$ is the number density of the collision partner,
and $\sigma$ is the cross section of the collision partner. Then, the timescale of the collision is $t \sim \frac{1}{n_{cl}\sigma v}$ where $v$ is the velocity of the cloud
with respect to the collision partner. After the calculation described in Appendix \ref{sec:collision}, we obtain
\begin{equation}
t \sim {\rm 1.6\,Myr}~ \left (\frac{R_{cl}}{\rm 20\,pc}\right )^{-2}  \left(\frac{M_{cl}}{1.5\times 10^5 M_{\odot}}\right) \left (\frac{v}{\rm 20\,km\,s^{-1}}\right )^{-1},
\end{equation}
where $R_{cl}$ and $M_{cl}$ are the radius and the mass of a giant molecular cloud (GMC), respectively.
Here we assume the mass and the radius of a GMC to be that of typical Galactic GMCs \citep{2009ApJ...699.1092H,2015ApJ...801...25L}.
This derived value of collision timescale is only a small fraction ($\sim 16\%$) of the orbit timescale. 
Those frequent inelastic collisions can reduce the velocity dispersion within the stream,
and the collision frequency gradually decreases. The change in collision frequency can reduce the fractional abundances of shock tracers
as discussed below.

\subsubsection{Timescale of chemical variation}
We find that fractional abundances of shock tracers vary gradually along the orbit. 
Here we discuss how this gradual change is caused considering the timescale of chemical and physical processes. 
The timescale of CH$_3$OH and SiO to freeze onto dust grains is $\sim 10^9/n$ yr, relatively short for moderately dense gas ($\lesssim 10^5$ yr for $n\gtrsim 10^4$ cm$^{-3}$) \citep{1999ApJ...523L.165C}. 
Because the orbital timescale of the circumnuclear ring is around 10 Myr, its size is around 500 pc in diameter,
and our angular resolution is in the order of 10 pc, we should observe no spatial shift between the location of shocks and the location of enhancement of shock tracers.
That is, shock tracers are only seen locally close to the location of shocks.
In other words, the observed gradual variation of chemical abundances suggests gradual variation in the frequency of shocks.

\subsubsection{Timescale of star formation}\label{sec:sftime}
From the fast variation of the chemistry of shock tracers, we consider that the collision timescale is relatively short $\sim 1$ Myr.
After a collision, the timescale of the star formation can be compared with the free fall time, $\tau_{ff} = {3\times 10^5} \left(\frac{n}{10^4 {\rm \,cm^{-3}}} \right)^{-1/2}$ yr.
The star formation usually takes place in the free-fall time or longer.
This timescale is shorter than the time it takes for the gas to move from N1 to N3 when $n \gtrsim 10^3$ cm$^{-3}$.
If the star formation at N3 is induced by a collision at N1, the collision must have compressed the gas to $n=10^3-10^4$ cm$^{-3}$.
We leave it to theoretical studies whether this level of compression is reasonable or not. 
Meanwhile, there are indeed Galactic star formation regions where collisions seem to have 
triggered star formation within $\lesssim1$ Myr \citep{2011ApJ...738...46T,2014ApJ...780...36F,2018ApJ...859..166F}.

Information on stellar ages at N2 and N3 can confirm the scenario of collision-induced star formation if they have ages around 2-3 Myr.
Age-dating of starburst has also been conducted for star clusters in and around the circumnuclear ring of M83. 
However, most young clusters are located in regions that are unobscured in optical or NIR wavelength,
and their ages are in the range of 5-10 Myr \citep{2001AJ....122.3046H,2008MNRAS.385.1110H,2010MNRAS.408..797K}.
Although \citet{2010MNRAS.408..797K} reported that there is a gradient of stellar ages in clusters near N2, N4, and N5 clouds
with ages of 5.5, 6.8, and 8.6 Myr, their observations likely did not trace stars formed at the orbit intersection on the northeastern stream
if they travel on the circumnuclear ring with the orbital period of 10 Myr. 
There are a few clusters with ages of a few Myr around the N2 cloud in the observations by \citet{2001AJ....122.3046H},
but there is no work reporting the ages of obscured star formation around the N2 and N3 clouds that we observe with radio recombination lines.
Therefore, these studies on stellar ages cannot be used to confirm the scenario of collision-induced star formation.

With the consideration of factors discussed in Sections \ref{sec:orbit} - \ref{sec:sftime}, 
we conclude it plausible that the star formation around N2 \& N3 was triggered by the collisions at the orbit intersection
if these collisions can compress the gas up to $n=10^3 - 10^4$ cm$^{-3}$ or higher.

\subsection{Comparison between the northeastern stream and the southwestern stream}
As explained in Section \ref{sec:molabun}, there is difference between the chemical composition in the northeastern and southwestern streams. 
The enhanced fractional abundance of methanol at S1 and S2 suggests that there are shocks at the orbit intersection in the southwestern stream, which is similar to the case in N1.
On the other hand, the SiO/$^{13}$CO intensity ratio is lower at S1 or S2 than at N1 and N2.  
Because SiO is known to be enhanced with strong shocks \citep[$> 25$ km s$^{-1}$][]{2008A&A...482..809G}, 
strong shocks must be more frequent in the northeastern stream.
On the southwestern stream, there is no sign of active star formation traced by the radio recombination line, either.

There are a few possible reasons for this lack of star formation in the southwestern stream. 
One possible reason is the lack of the gas supply.
As it is obvious from the moment 0 maps (e.g., $^{13}$CO($1-0$) map in Figure \ref{fig:mom0}), 
the distribution of the molecular gas is not uniform along the circumnuclear ring.
 If the gas inflow from the bar has been scarce in the southwestern stream in the past few Myr or so,
this lack of fresh gas supply suppresses star formation.
 If this is the case, the gas currently located at S1-S3 may result in star formation in a few Myr
as the clouds reach the intersection with the nuclear ring.
Another possible reason is the lack of strong shocks. Although CH$_3$OH fractional abundance is moderately enhanced at S1-S2 and the moment 2 map shows 
higher velocity dispersion at S1 compared with the upstream, there is less emission of SiO
in the southwestern stream. This trend indicates that strong shocks are less frequent there.
On the other hand, the velocity dispersion shown in the moment 2 map (Figure \ref{fig:mom1mom2} right) is slightly larger in the southwestern stream.
More turbulence on the southwestern stream can also restrain star formation.
 Although it seems contradictory to the less SiO intensity on the southwestern stream, the southwestern stream may have 
 more turbulence with low-velocity shocks.
There is yet another possible scenario where the collision-induced star formation is not the major mode of star formation.
\citet{2013ApJ...769..100S} proposed that collision-induced star formation does not provide as high star formation rate as
star formation due to self gravity within the circumnuclear ring when the gas inflow rate is large.
It is because the gas flows through the orbit intersection without colliding. However, this scenario is unlikely in this particular case 
of the southwestern stream of M83 because the gas is more scarce on this stream than on the northeastern stream.
The column densities of $^{13}$CO at S1 and S2 are about a half of the column densities at N1 and N2.

\subsection{Other contributing factors to the chemical composition}
In addition to the chemical processes discussed in our proposed scenario in Section \ref{sec:prop_scenario},
there are other factors contributing to the chemistry. First, we discuss the case of PDR tracers.
Although CCH and CN have been claimed to be PDR tracers, their fractional abundances depend 
not only on the UV-field strength, but also on the gas volume density because
reactions involving UV photons become more effective in lower-density environments.\footnote{Reactions involving UV photons, such as photodissociation or photoionization reactions, are one-body reactions,
 and their rates are proportional to the density $n$. Meanwhile, rates of two-body reactions are proportional to $n^2$.
 Therefore, two-body reactions become more effective than reactions involving UV-photons at higher densities.}
 Theoretical studies have shown that the quantity $G_0/n$ is the control parameter of PDR chemistry \citep[e.g., ][]{1996ApJ...468..269D}
 where $G_0$ is the UV-field strength.
 This dependence on the density has also been discussed in \citet{2015A&A...573A.116M} in the context of extragalactic chemical observations.
Our results show that the fractional abundances of CCH are almost constant among positions.
If the emission of CCH comes from low-density regions of molecular clouds, even clouds without star formation
can emit some amount of CCH. This may explain why the fractional abundances of CCH
do not necessarily correlate with the strengths of star formation.

Unlike PDR tracers, some species are more prone to dissociation by UV-photons. 
One example is methanol. We proposed that the enhancement of CH$_3$OH at N1 is caused by more frequent shocks
than at downstream positions. However, the decreased abundances of methanol at N2 or N3 may be caused by photodissociation
from the star formation there. On the other hand, the enhanced CH$_3$OH/$^{13}$CO at the entering point from the bar
is also seen on the southwestern stream, where there is no obvious star formation.
Therefore, it is likely that some methanol emission is still enhanced due to shocks.

Abundances of some species may also be affected by changes in the temperature of dust grains.
For example, when the dust grains are warm, the production rate of methanol is suppressed. 
Such heating of dust may be effective in active star-forming regions.
The N$_2$H$^+$ fractional abundances may also be affected by the dust temperature because CO may be less depleted on warm dust grains
and there is a major destruction route of N$_2$H$^+$ through CO \citep{2001ApJ...557..209B}.

\subsection{Comparison with other galactic centers}
There are galactic centers that show chemistry similar to the one in M83 and there are ones that do not.
Our proposed scenario is similar to what was suggested by \citet{2005ApJ...618..259M} for IC 342.
The enhancement of CH$_3$OH and N$_2$H$^+$ at the bar-ring orbit intersections was also seen in IC 342.

As mentioned earlier, NGC 1097 is an example where star formation seems independent from the galactic orbits.
\citet{2015A&A...573A.116M} studied the circumnuclear ring of NGC 1097 for the spatial abundance variation of such species as 
SiO, CCH, HCN, HCO$^+$, HNCO, CS, SO, and HC$_3$N.
Unlike our results in M83, there is no large variation of the chemical composition in the circumnuclear ring of NGC 1097 except for CCH.
This difference may be caused by the presence of collision-induced star formation in M83,
although we caution that a similar large scale chemical variation due to cloud collision may be present also in NGC 1097 
at a low level below the sensitivity of their cycle 0 observations.
In the case of NGC 1097, the scenario by \citet{2013ApJ...769..100S} may be relevant because of the high gas inflow rate.

\citet{2017ApJ...835...76M} analyzed the chemical composition of the Galactic central molecular zone (CMZ) using the survey data by \citet{2012MNRAS.419.2961J}.
They analyzed ratios of molecular line intensities with respect to the total column density of the molecular gas. 
Spatial variation of these ratios should reveal the variation of fractional abundances of those species provided that the excitation temperature is constant.
The interpretation of these data in relation with the galactic orbits is more difficult in the Galactic Center than in M83 because of the edge-on configuration of our Galaxy.
There are models of Galactic Center structure to explain the observed position-velocity diagram \citep[e.g., ][]{2004MNRAS.349.1167S,2015MNRAS.447.1059K}.
According to the model by \citet{2004MNRAS.349.1167S}, the $1.3^{\circ}$ cloud is at the orbit intersection from the bar to the circumnuclear ring. 
SiO is indeed enhanced in the $1.3^{\circ}$ cloud and the $1.6^{\circ}$ cloud as seen in \citet{2010A&A...523A..45R} from the SiO/H$^{13}$CO$^+$ intensity ratio.
In \citet{2017ApJ...835...76M}, the enhancement in the $1.3^{\circ}$ cloud and the $1.6^{\circ}$ cloud was not seen because their analysis only included regions above
a certain column density threshold, and these clouds have column densities below the threshold.
Therefore, the SiO fractional abundance is highest in Sgr B2 in \citet{2017ApJ...835...76M}.
Sgr B2 clouds were suggested to be under the influence of tidal compression in the model by \citet{2015MNRAS.447.1059K},
and frequent shocks may also be relevant there.
N$_2$H$^+$ is also most enhanced in Sgr B2 according to the map by \citet{2017ApJ...835...76M}.
Partly because of the edge-on configuration, we are unable to see similarity of the chemistry of the Galactic Center CMZ with that of M83
except for frequent shocks at the 1.3$^{\circ}$ cloud.

\section{Summary} \label{sec:sum}
We have presented multi-species, multi-transition ALMA observations in the galactic center region of M83. 
We have studied the variation of chemistry as the gas flows from the bar into the circumnuclear ring and moves to the star-forming regions.
Here are the summary of our results.
\begin{itemize}
\item Shock tracers (SiO and CH$_3$OH) are found to be enhanced at the two locations where the leading-edge gas lanes on the bar are connected to the circumnuclear ring.
\item In the northeastern stream, the N$_2$H$^+$ fractional abundance decreases as the gas travels further from the orbit intersection.
This variation in the fractional abundance indicates gas compression at the orbit intersection
if N$_2$H$^+$ emission preferentially comes from dense gas.
Signs of active star formation are seen from the radio recombination line H41$\alpha$ and CN further downstream.
This variation can be explained by the star formation induced by the collisions of molecular clouds at the orbit intersection and by the gas compression through the collisions.
\item If the star formation in the northeastern stream is indeed triggered at the orbit intersection, the timescale of star formation is 2-3 Myr.
The collision frequency at the orbit intersection is estimated to be on the order of $\sim 1$ Myr, and star formation within a few Myr is possible 
if the gas compression is effective enough to allow gravitational collapse in a short timescale on the order of 1 Myr.
\item The southwestern stream lacks the sign of gas compression and star formation. This difference from the northeastern stream may come from the lower amount of 
gas in the southwestern stream, or the lack of strong shocks.
\item The chemistry in the center of M83 and its dependence on the galactic structure is similar to what was observed in IC 342. 
On the other hand, the chemical variation of M83 shows little similarity with that in the circumnuclear ring of NGC 1097. 
\end{itemize}

We have shown here that astrochemistry can be used to study collision-induced star formation in nearby galaxies.
At the same time, it is still unclear what the exact cause is for presence or absence of collision-induced star formation.
Further astrochemical studies in a larger sample may help resolving the condition for the collision-induced star formation
and revealing more of the ISM evolution related to galactic gas dynamics.

\acknowledgments
We thank the anonymous referee for his/her constructive comments and
the ALMA staff for the service in observations, quality assessment, and technical help at the local ALMA Regional Center.
N.H. and S.K. acknowledge the financial support from Ministry of Science and Technology in Taiwan grant MOST 108-2112-M-001-015-.
D.R. acknowledges partial support by the Collaborative Research Council 956, 
subproject A5, funded by the Deutsche Forschungsgemeinschaft (DFG).
This paper makes use of the following ALMA data: ADS/JAO.ALMA\#2012.1.00762.S, \#2015.1.01177.S, and \#2016.1.00164.S. 
ALMA is a partnership of ESO (representing its member states), 
NSF (USA) and NINS (Japan), together with NRC (Canada), MOST and ASIAA (Taiwan), and KASI (Republic of Korea),
 in cooperation with the Republic of Chile. The Joint ALMA Observatory is operated by ESO, AUI/NRAO and NAOJ.
 This research has made use of the NASA/IPAC Extragalactic Database (NED) which is operated by the Jet Propulsion Laboratory, 
 California Institute of Technology, under contract with the National Aeronautics and Space Administration.
 This research made use of APLpy, an open-source plotting package for Python \citep{2012ascl.soft08017R}.

 {\it Facilities: ALMA}

{\it Software: CASA \citep[v4.7, v4.7.2., v5.4][]{2007ASPC..376..127M}, APLpy \citep{2012ascl.soft08017R}, MADCUBA (Mart\'in et al. 2019, submitted)}


 \begin{figure}
\includegraphics[width=0.45\textwidth,trim = 0 0 0 0]{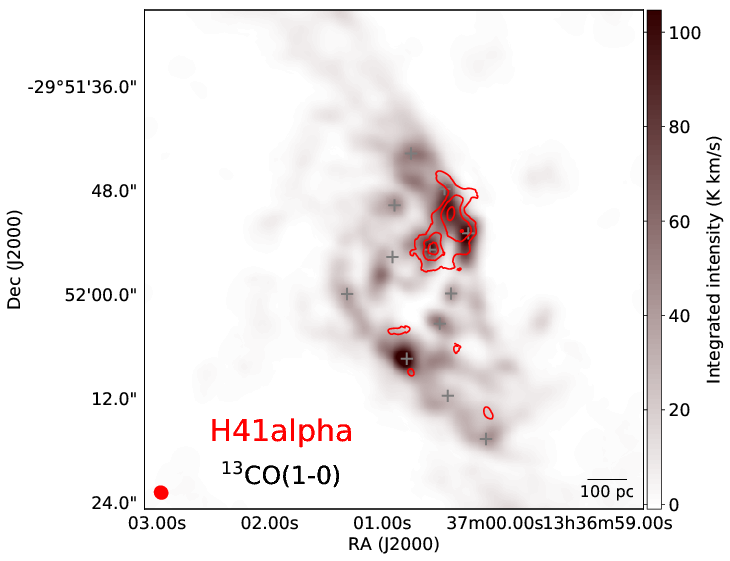}
\caption{Velocity-integrated intensity map of H41$\alpha$ (red contours) overlaid on the velocity-integrated intensity map of $^{13}$CO(1-0) (grey scale).
The contour levels of H41$\alpha$ are every 0.96 K km s$^{-1}$ with the typical rms of 0.32 K km s$^{-1}$ (see details of moment 0 maps in Appendix \ref{sec:app_mom0}).\label{fig:mom0_overlay}}
\end{figure}

%






\appendix
\section{Velocity-integrated intensity maps}\label{sec:app_mom0}
\begin{figure*}
\centering{
\includegraphics[width=0.3\textwidth,trim = 0 0 0 0]{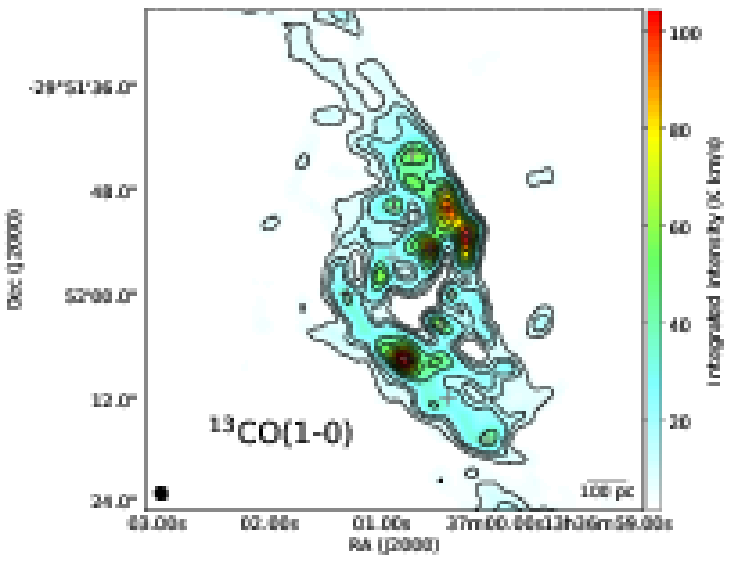}
\includegraphics[width=0.3\textwidth,trim = 0 0 0 0]{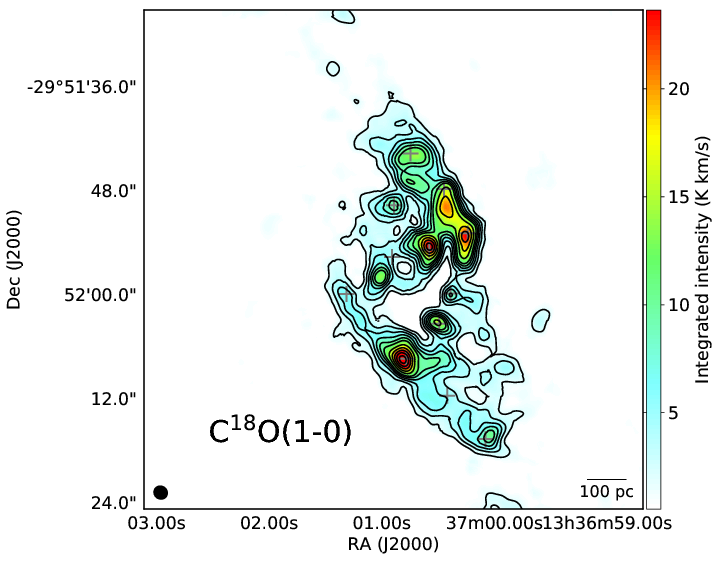}
\includegraphics[width=0.3\textwidth,trim = 0 0 0 0]{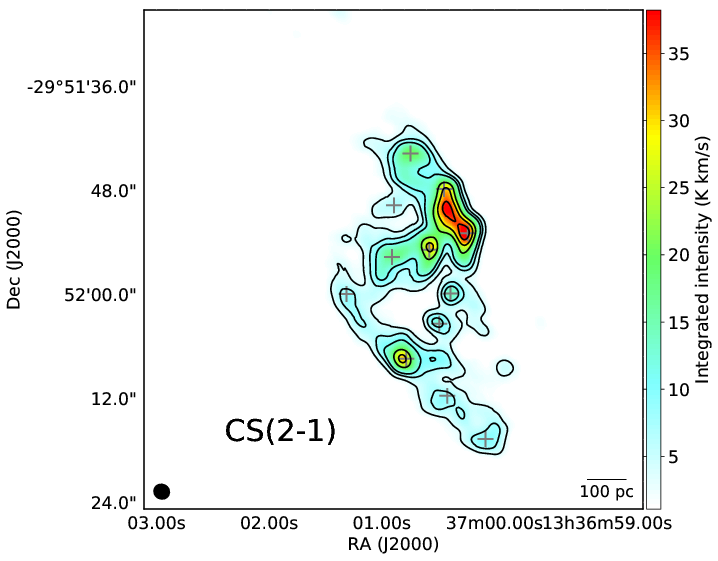}
}
\centering{
\includegraphics[width=0.3\textwidth,trim = 0 0 0 0]{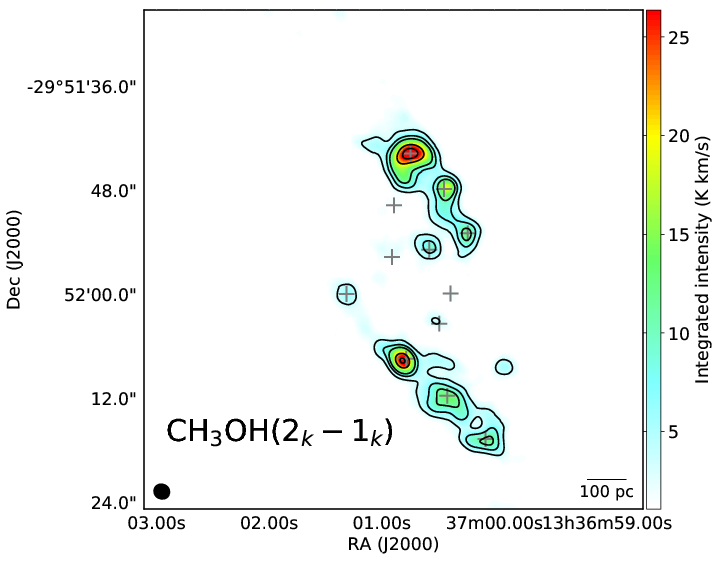}
\includegraphics[width=0.3\textwidth,trim = 0 0 0 0]{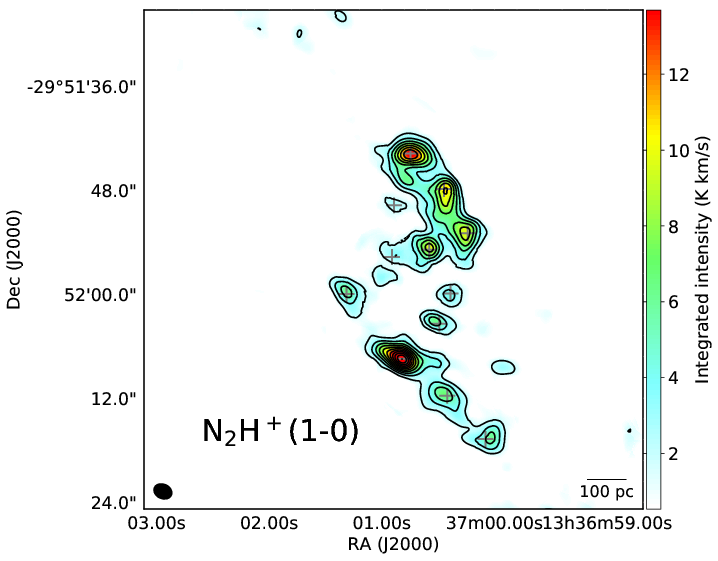}
\includegraphics[width=0.3\textwidth,trim = 0 0 0 0]{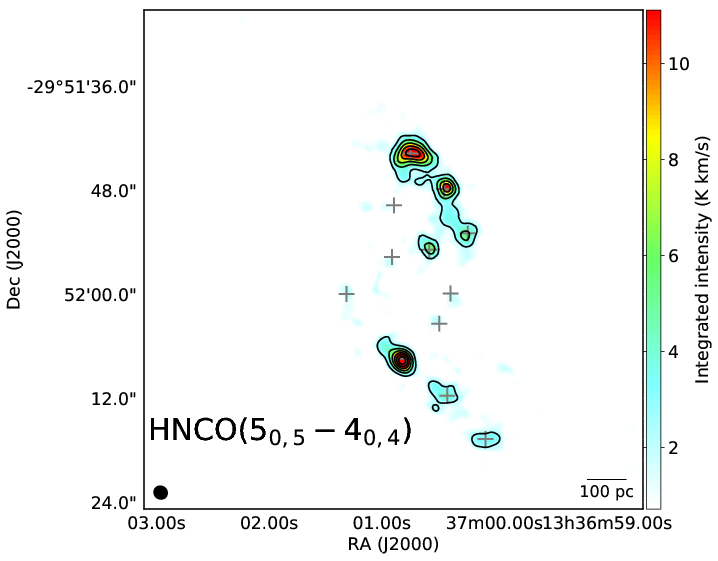}
}
\centering{
\includegraphics[width=0.3\textwidth,trim = 0 0 0 0]{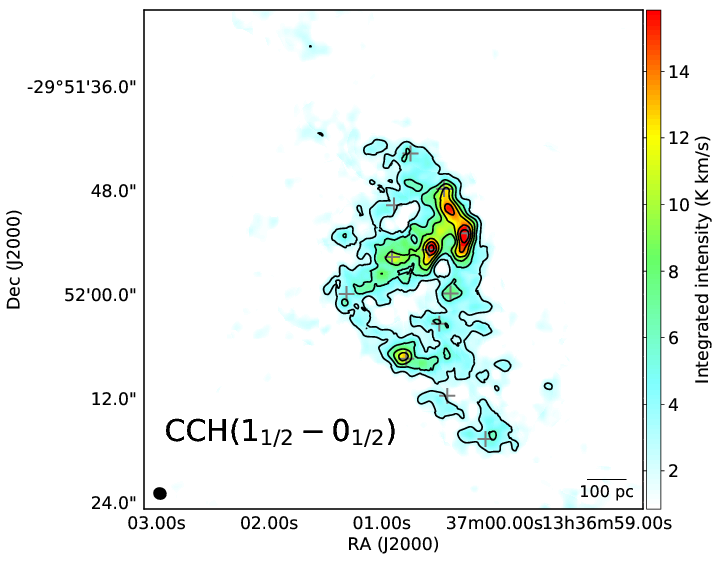}
\includegraphics[width=0.3\textwidth,trim = 0 0 0 0]{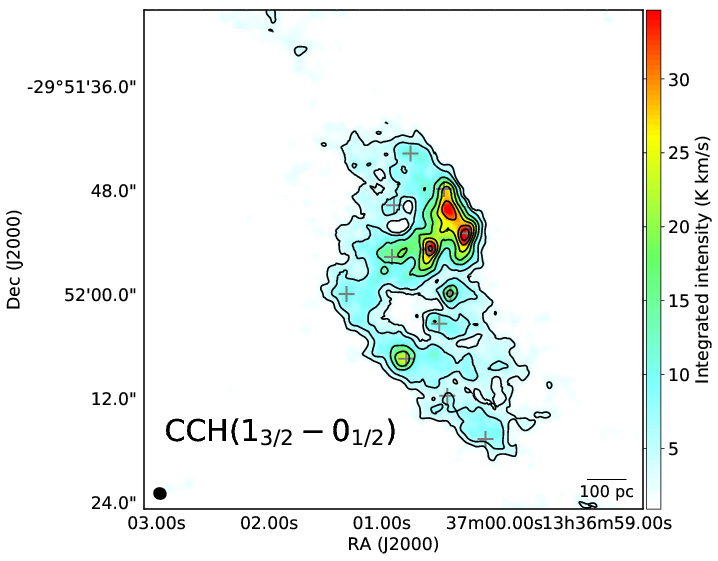}
\includegraphics[width=0.3\textwidth,trim = 0 0 0 0]{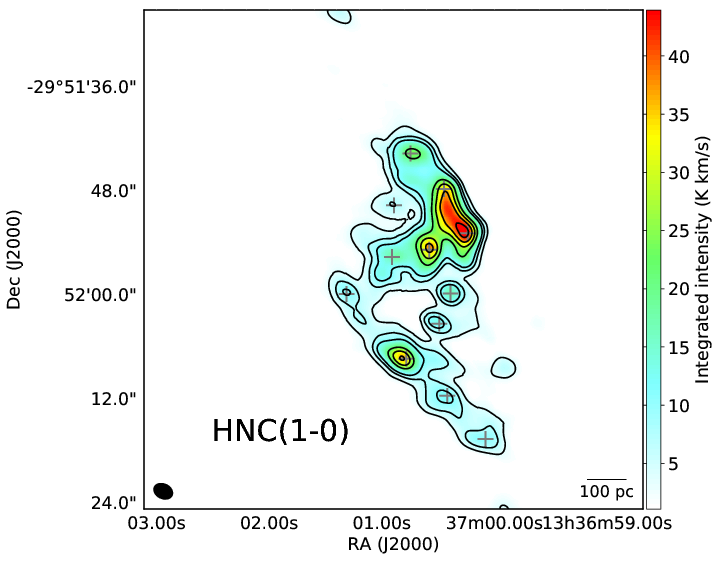}
}

\centering{
\includegraphics[width=0.3\textwidth,trim = 0 0 0 0]{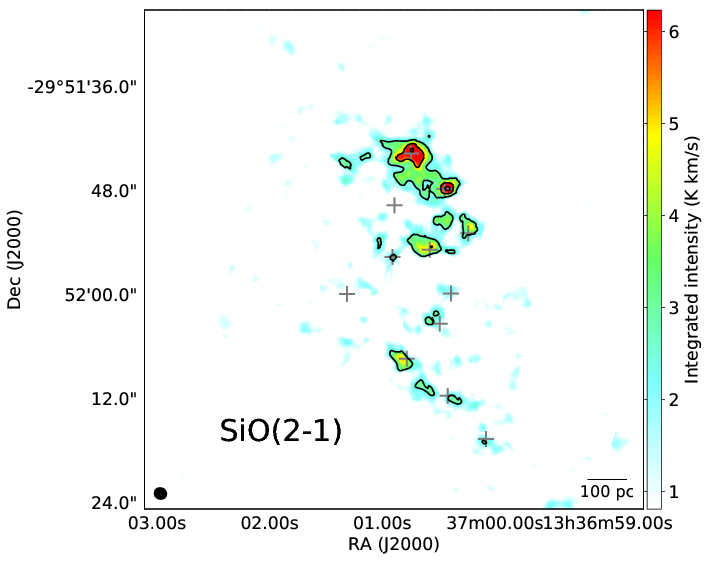}
\includegraphics[width=0.3\textwidth,trim = 0 0 0 0]{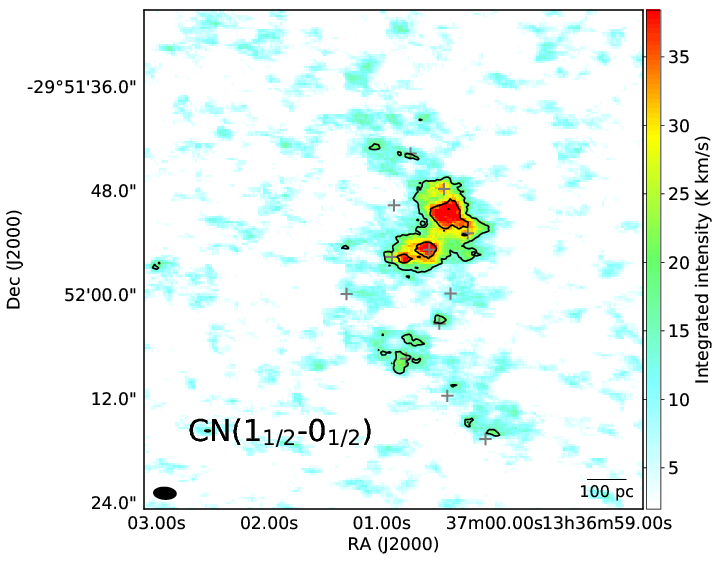}
\includegraphics[width=0.3\textwidth,trim = 0 0 0 0]{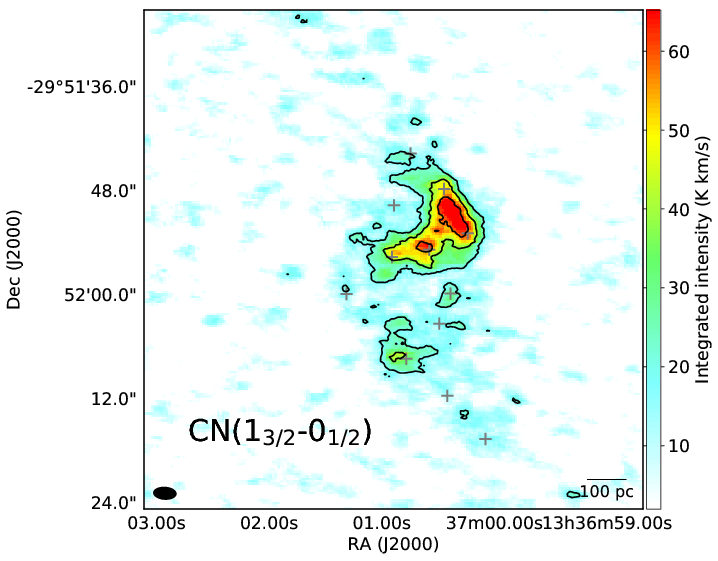}
}
\caption{Velocity-integrated intensity maps of some of the detected lines before the primary beam correction. 
The analyzed positions N1-5, S1-5, and C1-2 are shown as grey plus signs.
The synthesized beam is shown as a black ellipse on the lower left corner.
Their beam sizes are listed in Table \ref{tab:im_param}.
The noise level spatially varies in each map because the number of integrated channels varies with position
due to our masking. The noise at the N1 position is listed below as $\sigma_{\rm N1}$.
See Appendix \ref{sec:app_mom0} to obtain the noise at other positions.
Contour levels are plotted at 5.9, 11.8, 17.6, and every 11.8 K kms$^{-1}$ for $^{13}$CO($1-0$) with a typical noise of $\sigma_{\rm N1} = 0.56$ K km s$^{-1}$, 
1.7, 3.5, 5.2, 7.0, 8.7, and every 3.5 K km s$^{-1}$ for C$^{18}$O($1-0$) with $\sigma_{\rm N1} = 0.44$ K km s$^{-1}$, 
3.4, 6.7, 10.1, and every 6.7 K km s$^{-1}$ for CS($2-1$) with $\sigma_{\rm N1} = 0.34$ K km s$^{-1}$,  
3.2, 6.5, 9.7, and every 6.5 K kms$^{-1}$ for CH$_3$OH($2_k-1_k$) with $\sigma_{\rm N1} = 0.33$ K km s$^{-1}$,
every 2.1 K kms$^{-1}$ for N$_2$H$^+$($1-0$) with $\sigma_{\rm N1} = 0.39$ K km s$^{-1}$, 
every 2.9 K km s$^{-1}$ for HNCO($5_{0,5}-4_{0,4}$) with $\sigma_{\rm N1} = 0.42$ K km s$^{-1}$, 
3.3, 6.6, 9.9, and every 6.6 K km s$^{-1}$ for  CCH($1_{3/2}-0_{1/2}$) with $\sigma_{\rm N1} = 1.1$ K km s$^{-1}$,
every 2.9 K km s$^{-1}$ for CCH($1_{1/2}-0_{1/2}$) with $\sigma_{\rm N1} = 0.95$ K km s$^{-1}$, 
4.0, 7.9, 11.9, and every 7.9 for HNC($1-0$) with $\sigma_{\rm N1} = 0.37$ K km s$^{-1}$, 
every 2.7 K km s$^{-1}$ for SiO($2-1$) with $\sigma_{\rm N1} = 0.9$ K km s$^{-1}$.
For CN lines in the Band 3, we did not use the mask, and contour levels are 
every 16.7 K km s$^{-1}$ for CN(1$_{1/2}$-0$_{1/2}$) with $\sigma = 6.0$ K km s$^{-1}$, 
every 19.8 K km s$^{-1}$ for CN(1$_{3/2}$-0$_{1/2}$) with $\sigma = 6.5$ K km s$^{-1}$.
\label{fig:mom0}}
\end{figure*}

\begin{figure*}
\centering{
\includegraphics[width=0.3\textwidth,trim = 0 0 0 0]{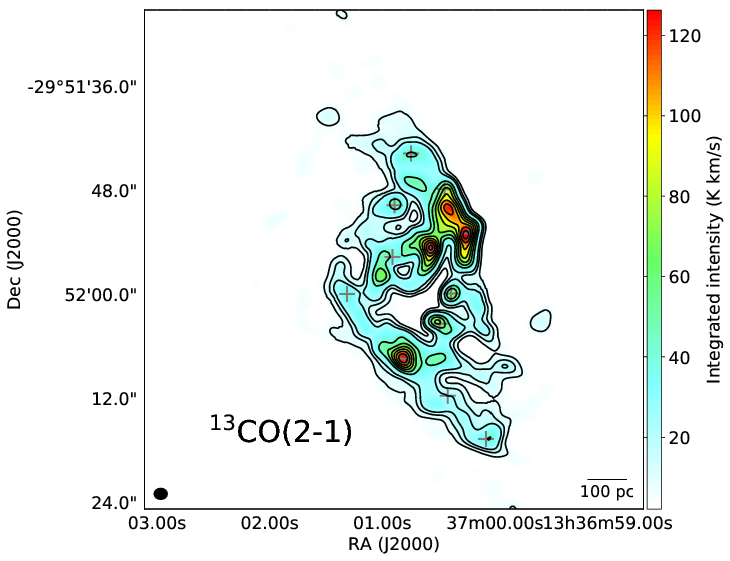}
\includegraphics[width=0.3\textwidth,trim = 0 0 0 0]{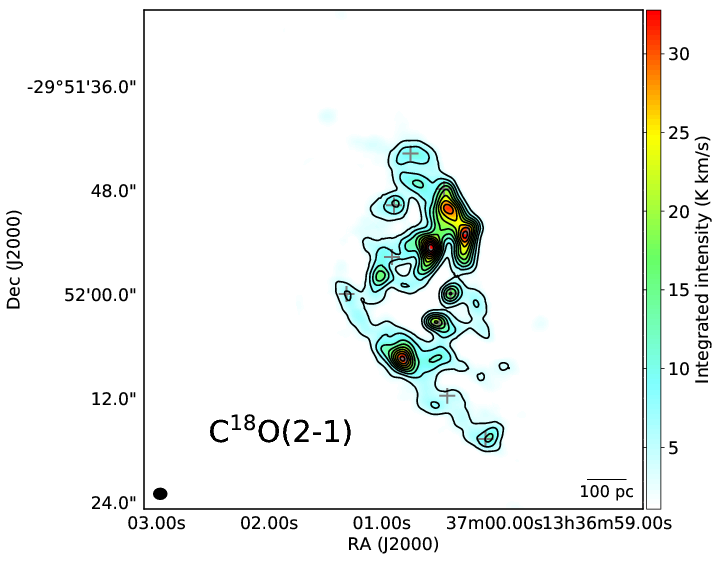}
\includegraphics[width=0.3\textwidth,trim = 0 0 0 0]{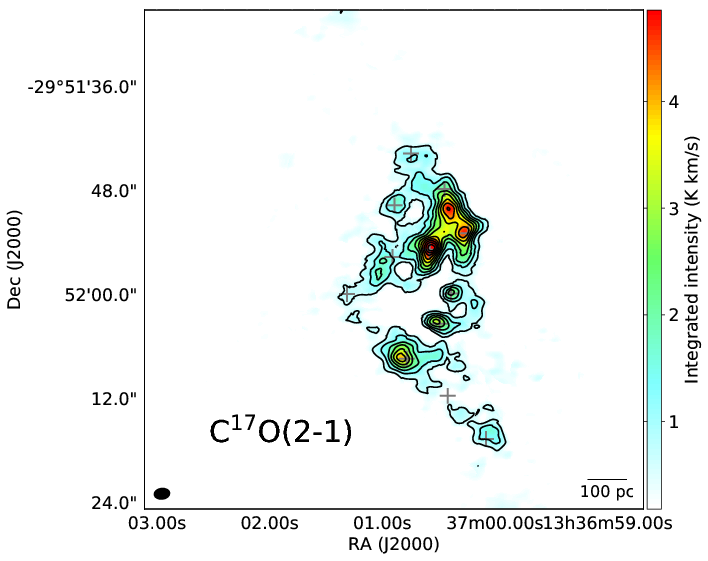}
}
\centering{
\includegraphics[width=0.3\textwidth,trim = 0 0 0 0]{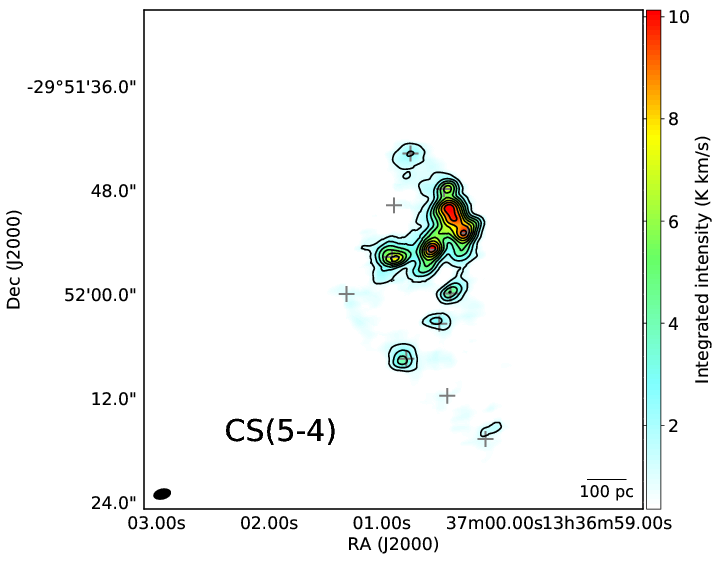}
\includegraphics[width=0.3\textwidth,trim = 0 0 0 0]{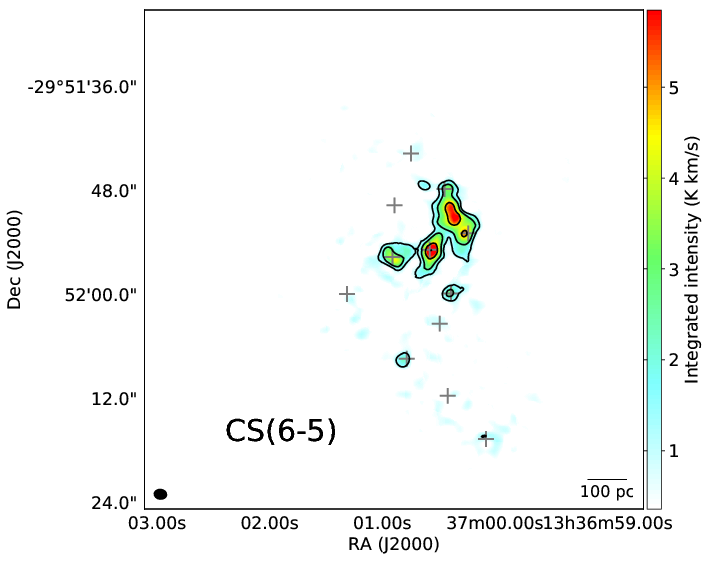}
\includegraphics[width=0.3\textwidth,trim = 0 0 0 0]{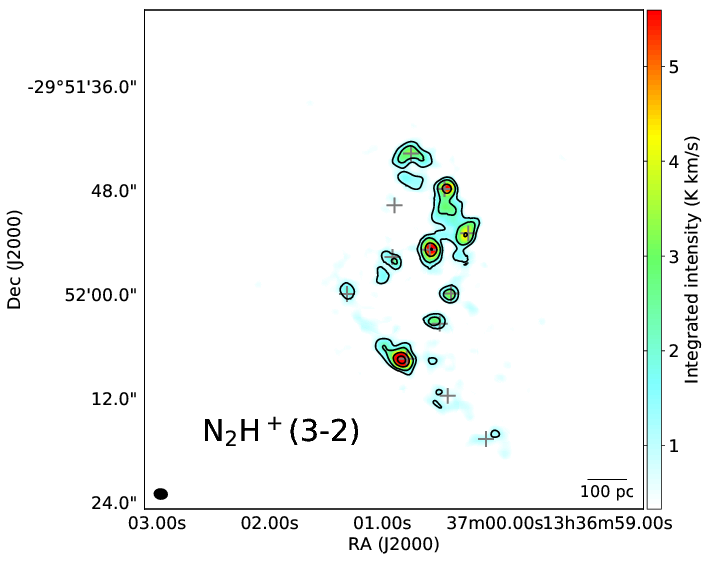}
}
\centering{
\includegraphics[width=0.3\textwidth,trim = 0 0 0 0]{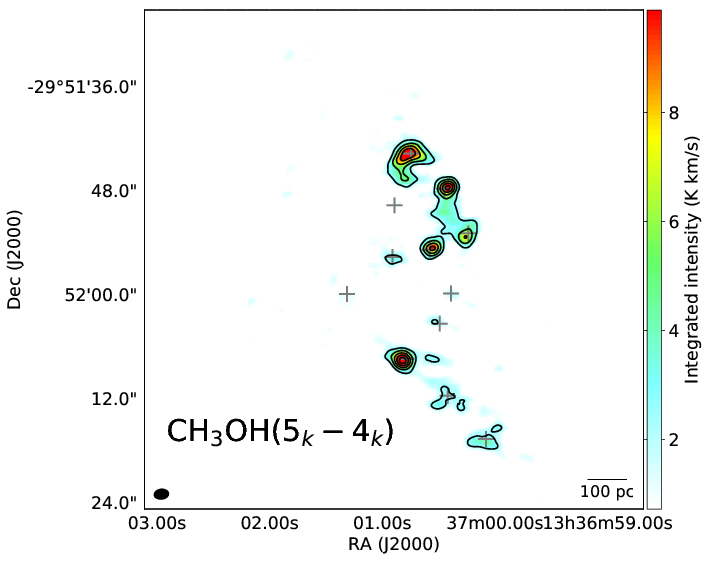}
\includegraphics[width=0.3\textwidth,trim = 0 0 0 0]{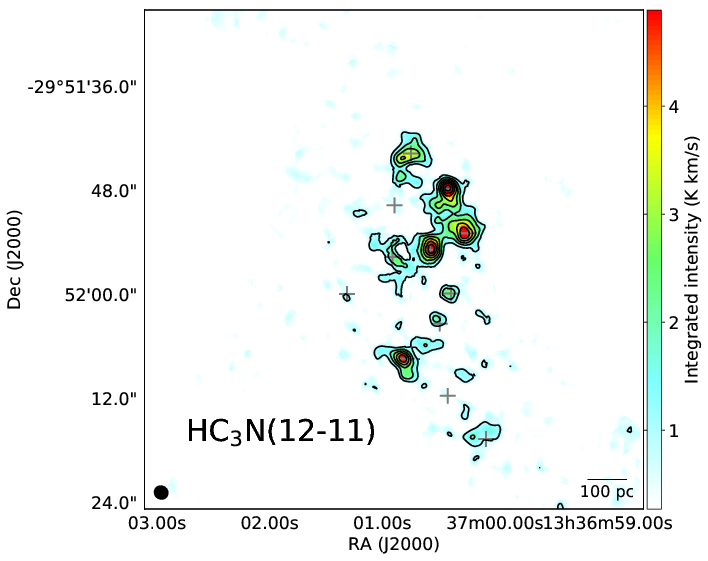}
\includegraphics[width=0.3\textwidth,trim = 0 0 0 0]{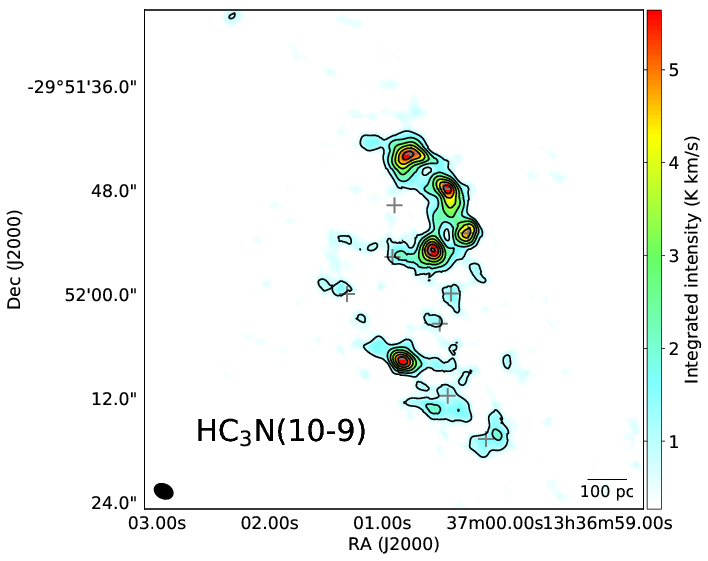}
}
\caption{Same as Figure \ref{fig:mom0}, but for other lines. 
Contour levels are plotted at 7.6, 15.1, 22.7, and every 15.1 K km s$^{-1}$ for $^{13}$CO($2-1$) with $\sigma_{\rm N1} = 0.53$ K km s$^{-1}$, 
every 3.6  K km s$^{-1}$ for C$^{18}$O($2-1$) with $\sigma_{\rm N1} = 0.31$ K km s$^{-1}$,
 every 0.6 K km s$^{-1}$ for C$^{17}$O($2-1$) with $\sigma_{\rm N1} = 0.20$ K km s$^{-1}$, 
 every 1.4 K km s$^{-1}$ for CS($5-4$) with $\sigma_{\rm N1} = 0.19$ K km s$^{-1}$,
1.2, 2.3, 3.5, and every 2.3 K km s$^{-1}$ for CS($6-5$) with $\sigma_{\rm N1} = 0.26$ K km s$^{-1}$, 
1.1, 2.2, 3.3, and every 2.2 K km s$^{-1}$ for N$_2$H$^+$($3-2$) with $\sigma_{\rm N1} = 0.37$ K km s$^{-1}$,
every 2.2 K km s$^{-1}$ for CH$_3$OH($5_k-4_k$) with $\sigma_{\rm N1} = 0.27$ K km s$^{-1}$,
 every 0.8 K km s$^{-1}$ for HC$_3$N($12-11$) with $\sigma_{\rm N1} = 0.41$ K km s$^{-1}$, 
 and every 0.8 K km s$^{-1}$ for HC$_3$N($10-9$) with $\sigma_{\rm N1} = 0.34$ K km s$^{-1}$.
\label{fig:mom0-2}}
\end{figure*}

\begin{figure*}
\centering{
\includegraphics[width=0.3\textwidth,trim = 0 0 0 0]{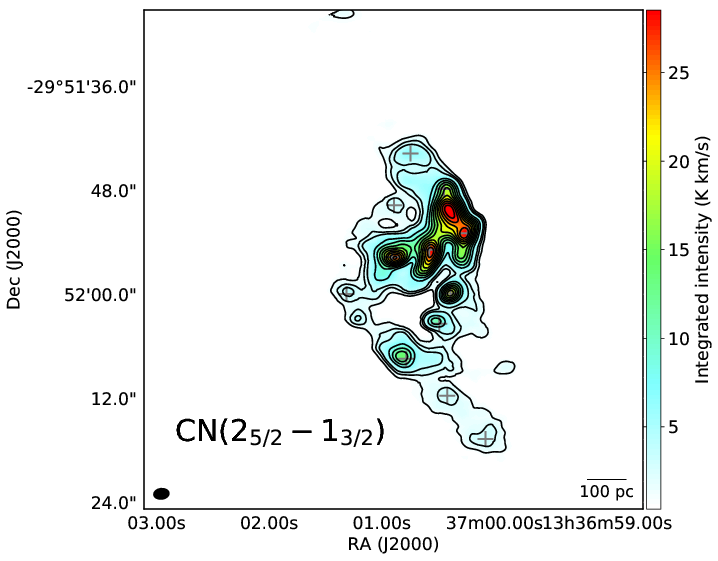}
\includegraphics[width=0.3\textwidth,trim = 0 0 0 0]{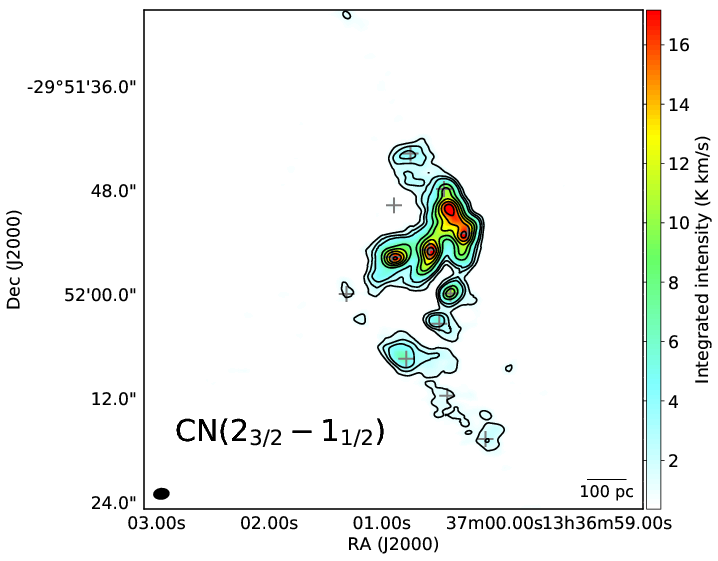}
\includegraphics[width=0.3\textwidth,trim = 0 0 0 0]{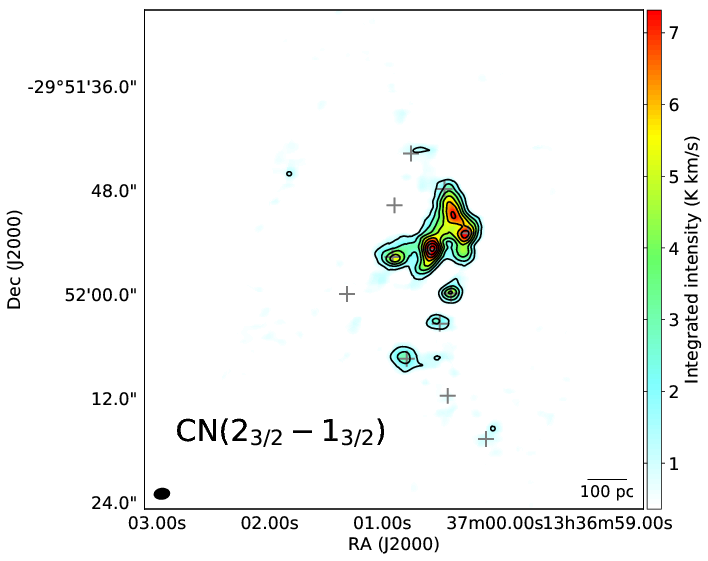}
}
\centering{
\includegraphics[width=0.3\textwidth,trim = 0 0 0 0]{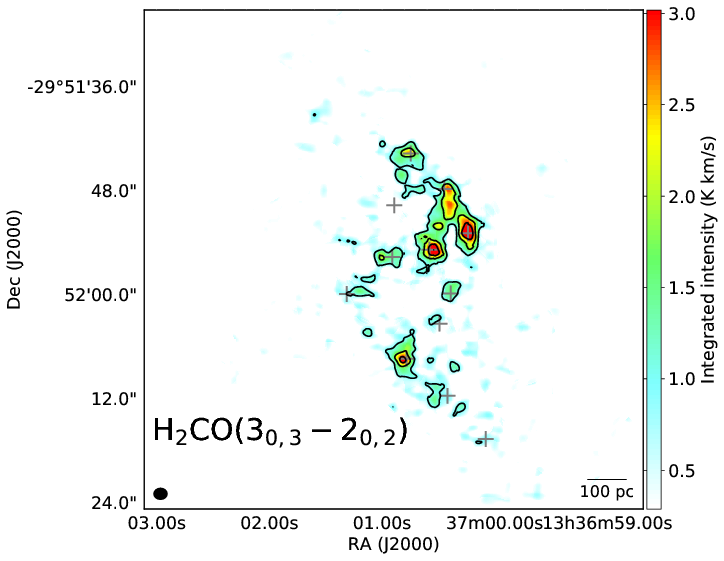}
\includegraphics[width=0.3\textwidth,trim = 0 0 0 0]{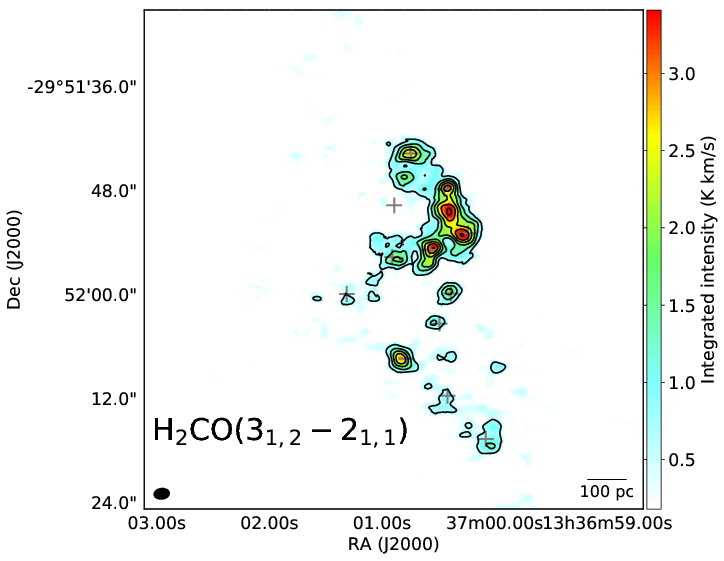}
\includegraphics[width=0.3\textwidth,trim = 0 0 0 0]{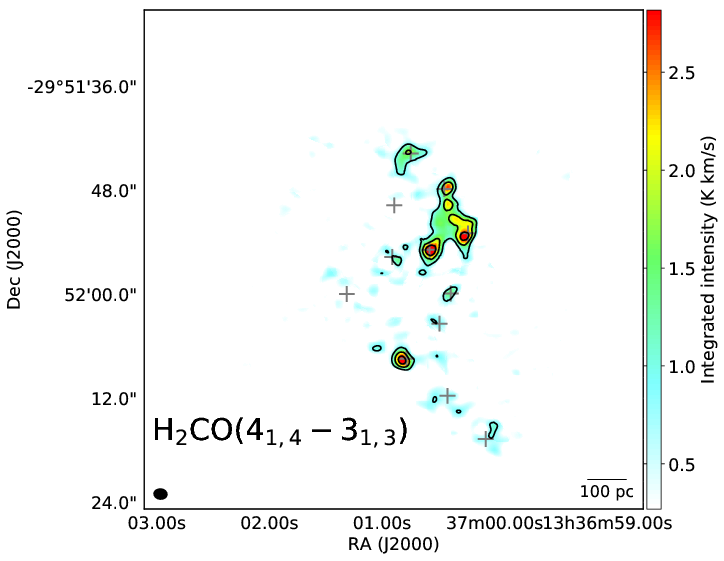}
}
\centering{
\includegraphics[width=0.3\textwidth,trim = 0 0 0 0]{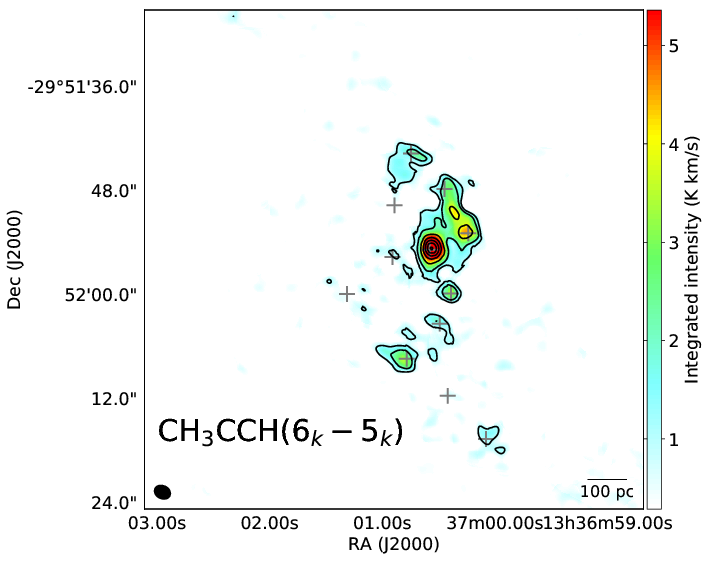}
\includegraphics[width=0.3\textwidth,trim = 0 0 0 0]{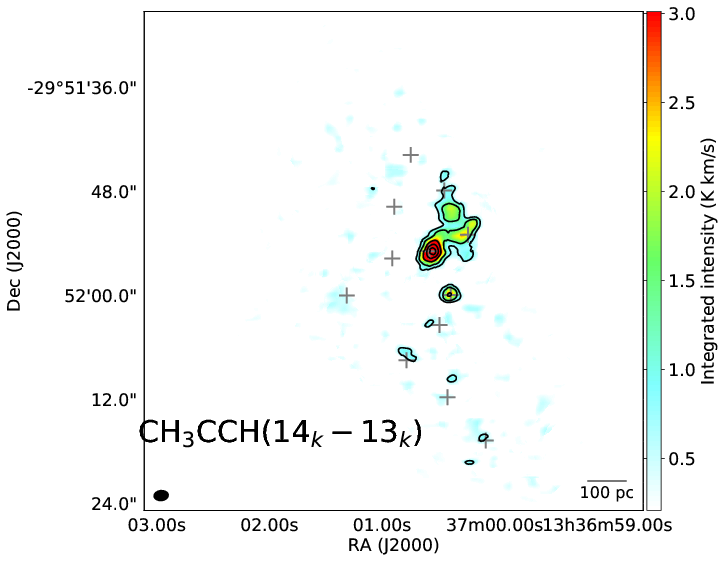}
\includegraphics[width=0.3\textwidth,trim = 0 0 0 0]{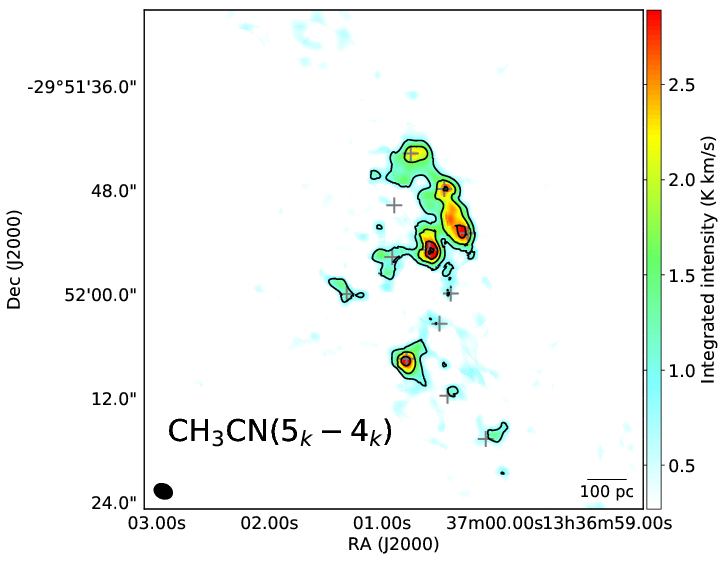}
}
\centering{
\includegraphics[width=0.3\textwidth,trim = 0 0 0 0]{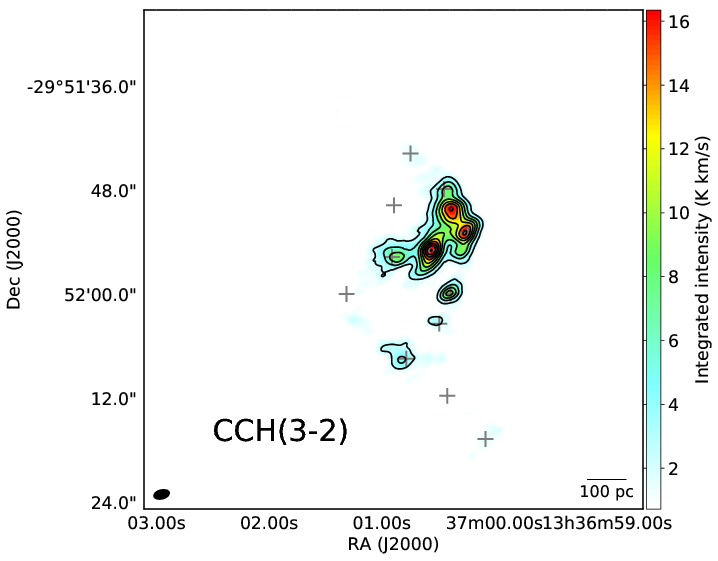}
\includegraphics[width=0.3\textwidth,trim = 0 0 0 0]{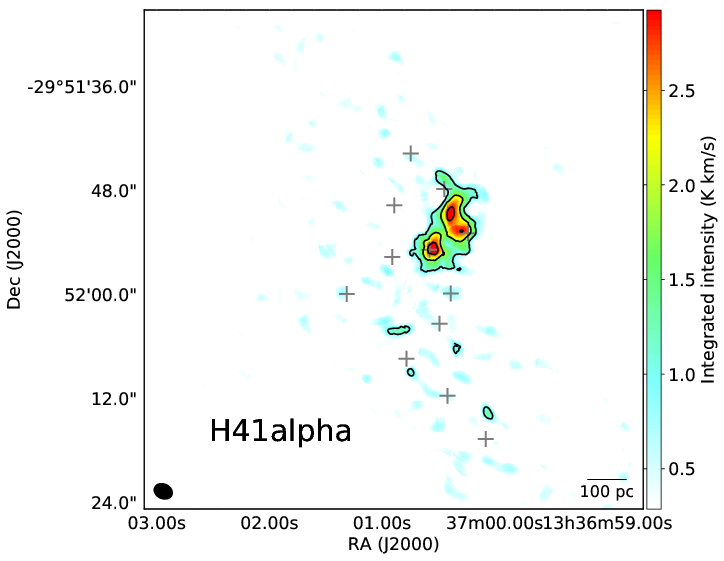}
}
\caption{Same as Figure \ref{fig:mom0}, but for other lines. 
Contour levels are plotted at 1.6, 3.1, 4.7, and every 3.1 K km s$^{-1}$ for CN($2_{5/2}-1_{3/2}$) with $\sigma_{\rm N1} = 0.16$ K km s$^{-1}$, 
1.8, 3.6, 5.4, and every 3.6 K km s$^{-1}$ for CN($2_{3/2}-1_{1/2}$) with $\sigma_{\rm N1} = 0.18$ K km s$^{-1}$,
 and every 1.7 K km s$^{-1}$ for CN($2_{3/2}-1_{3/2}$) with $\sigma_{\rm N1} = 0.18$ K km s$^{-1}$,
every 1.0 K km s$^{-1}$ for H$_2$CO($3_{0,3}-2_{0,2}$) with $\sigma_{\rm N1} = 0.32$ K km s$^{-1}$, 
every 0.6 K km s$^{-1}$ for H$_2$CO($3_{1,2}-2_{1,1}$) with $\sigma_{\rm N1} = 0.20$ K km s$^{-1}$,
every 1.0 K km s$^{-1}$ for H$_2$CO($4_{1,4}-3_{1,3}$) with $\sigma_{\rm N1} = 0.22$ K km s$^{-1}$, 
0.96, 1.9, 2.8, and every 1.9 K km s$^{-1}$ for CH$_3$CCH($6_k-5_k$) with $\sigma_{\rm N1} = 0.32$ K km s$^{-1}$,
0.7, 1.4, 2.1, and every 1.4 K km s$^{-1}$ for CH$_3$CCH($14_k-13_k$) with $\sigma_{\rm N1} = 0.23$ K km s$^{-1}$, 
every 0.8 K km s$^{-1}$ for CH$_3$CN($5_k-4_k$) with $\sigma_{\rm N1} = 0.34$ K km s$^{-1}$,
every 2.3 K km s$^{-1}$ for CCH($3-2$) with $\sigma_{\rm N1} = 0.22$ K km s$^{-1}$,  
and every 0.96 K km s$^{-1}$ for  H41$\alpha$ with $\sigma_{\rm N1} = 0.32$ K km s$^{-1}$.
 \label{fig:mom0-3}}
\end{figure*}

Figures \ref{fig:mom0}-\ref{fig:mom0-3} show velocity-integrated intensity maps in our observations.
To create moment 0 maps with as high signal-to-noise ratios as possible, 
we masked out the region with $^{13}$CO emission below 5 $\sigma$ in each velocity channel.
 This mask was applied to line cubes of all isolated lines,
and velocity-integrated intensity maps were created with the task immoments.
A number of species show multiple transitions contributing to the observed line profile (e.g. the hyperfine structure of CN or CCH)
 For those lines, we made masks with velocity shifts corresponding to those transitions. 
 The intersection of those masks was applied to the cube as a new mask.
 Because of the use of masks, the noise levels are not constant within the map.
 The error in the velocity-integrated maps can be estimated as $\sigma_{\rm integ} = \sigma_{1ch} \Delta v_{ch} \sqrt{N}$
 where $N$ is the number of channels integrated and $v_{ch}=10$ km s$^{-1}$ is the width of one channel.
 If there is no blending of multiple transitions, the number of channels integrated over at analyzed positions are 
N1 : 9, N2 : 12, N3 : 12, N4 : 8, N5 : 6, C1 : 17, C2 : 13, S1 : 10, S2 : 11, S3 : 11, S4 : 9.6, S5 : 7.
 To obtain the number of channels used for the images with multiple transitions integrated together, 
 add 10 to those numbers in the isolated line case for CN ($N=2-1,J=3/2-3/2$), 9 for 
 CN ($N=2-1,J=3/2-1/2$), 2 for CN (N=2-1,J=5/2-3/2), 
4 for CCH($N=1-0,J=3/2-1/2$),
2 for CCH ($N=1-0,J=1/2-1/2$),
7 for CCH ($N=3-2$),
and 22 for CH$_3$OH ($5_k-4_k$). 

\section{Continuum emission}

\begin{figure*}
\centering{
\includegraphics[width=0.3\textwidth,trim = 0 0 0 0]{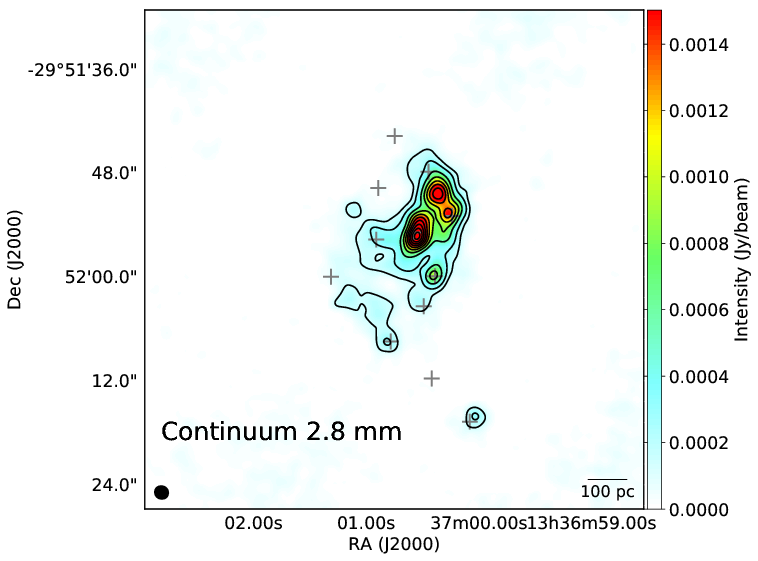}
\includegraphics[width=0.3\textwidth,trim = 0 0 0 0]{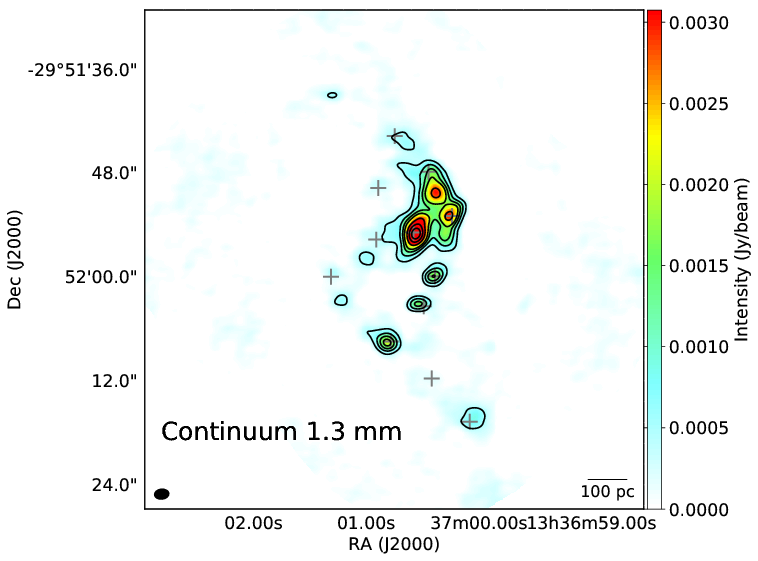}
\includegraphics[width=0.3\textwidth,trim = 0 0 0 0]{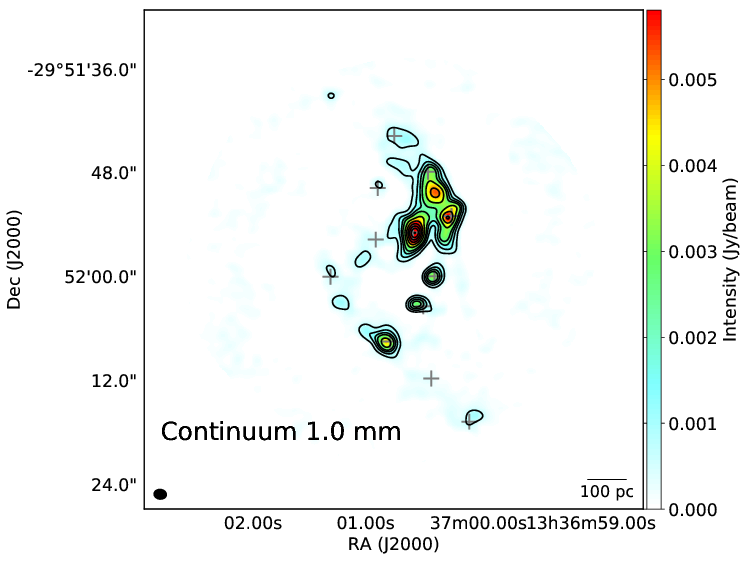}
}
\caption{Continuum images of the central region of M83 at 2.8, 1.3, 1.0 mm. Images are produced using line-free channels of one sideband (3.65-GHz wide). 
The contour levels are 6, 12, 18, 24, and every 12 $\sigma$ for 2.8mm ($1\sigma = 25\,\mu$Jy beam$^{-1}$), 6, 12, 18, 24, 30, 36, and every 12 $\sigma$ for 1.3 mm ($1\sigma = 77\,\mu$Jy beam$^{-1}$) , and 6, 12, 18, 24, and every 12 $\sigma$ for 1.0 mm ($1\sigma = 100\,\mu$Jy beam$^{-1}$). The synthesized beams are shown as black ellipses on the lower left corner, and their sizes are $1.8'' \times 1.7''$ for 2.8 mm, $1.8'' \times 1.3''$ for 1.3 mm, $1.6'' \times 1.3''$ for 1 mm. \label{fig:cont}}
\end{figure*}

Figure \ref{fig:cont} shows the maps of continuum emission at 2.8, 1.3, and 1.0 mm. The areas we detected the continuum emission
are smaller than those for major molecular lines such as CO isotopologues.
There is also visible difference between the continuum emission of 2.8 mm and those at 1.3 and 1.0 mm. 
This is most likely because the former has a sizable contribution from free-free and synchrotron emission while the latter is dominated 
by dust emission.

We estimate the contribution of free-free emission to the continuum emission in Band 3 from the flux of the radio recombination line H$41\alpha$,
which turns out to be 45-100 \%. 
We used a formula from \citet{1978ARA&A..16..445B} to estimate the free-free continuum intensity from the observed H$41\alpha$ intensity,
\begin{equation}
\int T_L dv/T_c = 6.76\times 10^3 \nu^{1.1} T_e^{-1.15},
\end{equation}
 where $\int T_L dv$ is the velocity-integrated intensity of a radio recombination line, $T_c$ is the intensity of continuum from free-free emission,
 $\nu$ is the frequency in the unit of GHz, and $T_e$ is the electron temperature.
Within the area of H41$\alpha$ emission, the averaged H41$\alpha$ velocity-integrated intensity is 0.94 K km s$^{-1}$, and the continuum intensity is 0.029 K. 
We obtained the fractional contribution of the free-free emission using the range of electron temperatures $T_e = 4000-10^4$ K from values in Galactic sources or in a starburst galaxy \citep{1983MNRAS.204...53S,2004MNRAS.347..237P,2015MNRAS.450L..80B}.
Presence of significant free-free emission explains different spatial distribution of continuum emission
between Band 3 and Bands 6 and 7. The continuum intensities are significantly larger in the higher-frequency bands, 
which must be due to dust emission.

\section{Estimate of collision frequency}\label{sec:collision}
The collision time scale is estimated as $t \sim \frac{1}{n_{cl}\sigma v}$ where $v$ is the velocity of the cloud
with respect to the collision partner, and $n_{cl}$ is the number density of clouds.
Meanwhile, the volume filling factor $\Phi$ can be expressed as $n_{cl} V_{cl}$ where $V_{cl}$ is the volume of a single molecular cloud.
If we assume a spherical cloud with a radius $R_{cl}$, the collision time scale can be expressed as $t \sim \frac{4}{3} \frac{R_{cl}}{\Phi v}.$
Here we estimate the volume filling factor as follows.
We assume the stream is flowing in a cylindrical structure with a radius of $r$ and a length of $l$.
Consider the number of molecular clouds within this cylinder to be $N_{cl}$, and the total volume of the cylinder to be $V_{tot}$.
Then, $\Phi = \frac{N_{cl}V_{cl}}{V_{tot}}.$ If we define $M_{tot}$ as the total gas mass in the volume and $M_{cl}$ as the mass of a single cloud,
we can express $N_{cl}$ as $N_{cl}=\frac{M_{tot}}{M_{cl}}$.
$M_{tot}$ can also be expressed as $M_{tot} = m_{H2}N_{H2}A$ where $A=2rl$ is the area of the cylinder seen from the side,
$m_{H2}$ is the mass of one hydrogen molecule,
and $N_{H2}$ is the averaged ${\rm H_{2}}$ column density of the cylinder seen from the side. 
Because $V_{cl} = \frac{4}{3}\pi R_{cl}^3$ and $V_{tot} = \pi r^2 l$,
we can rewrite the volume filling factor as $\Phi = \frac{8}{3} \frac{m_{H2} N_{H2} R_{cl}^3}{M_{cl} r}$.
From our data, the average column density of $^{13}$CO over $r=55$ pc ($2.5''$) is $N_{13CO} \sim 2\times 10^{16}$ cm$^{-2}$.
Assuming the $N_{13CO} = 10^{-6} N_{H2}$, $R_{cl} = 20$ pc, and $M_{cl}= 1.5 \times 10^5$ $M_{\odot}$, we obtain
$\Phi = 0.77 \left(\frac{R_{cl}}{20 pc}\right)^3 \left(\frac{M_{cl}}{1.5\times 10^5 M_{\odot}}\right)^{-1} \left(\frac{N_{H2}}{2\times 10^{22} {\rm cm^{-3}}}\right) \left(\frac{r}{55 pc}\right)^{-1}$ .




\bibliography{m83}

\begin{deluxetable}{cccccccccc}
\tablecolumns{10}
\tablewidth{0pc}
 \tabletypesize{\scriptsize}
\tablecaption{Observational Parameters\label{tab:obs_param}.}
\tablehead{\colhead{LSB Range} &\colhead{USB Range} & \colhead{Project code} &\colhead{Array}  &\colhead{Config}  &\colhead{Observation Date}  &\colhead{$N_{\rm ant}$}  &\colhead{Baseline} &\colhead{UV range}&\colhead {T$_{\rm on}$}\\
\colhead{(GHz)} &\colhead{(GHz)} &\colhead{} &\colhead{}  &\colhead{}  &\colhead{}  &\colhead{}  &\colhead{Range (m)} &\colhead{(k$\lambda$)} &\colhead{(min)}\\}
\startdata
  \hline
 85.61-89.34&97.54-101.27 &2015.1.01177.S &12m1 &C40-6&2016-09-20&37&14.4-3052&4.5-951.2&40\\
  &&&12m1 &C40-6&2016-09-22&37&14.3-3126&4.5-974.3&40\\
  &&&12m2 &C36-2/3&2016-04-18&39&14.5-594.6&4.5-185.3&40\\
    89.56 - 93.23&101.56-105.17 &2016.1.00164.S&12m&C40-3&2016-12-15&43&14.3-430.6&4.6-139.8&33\\
  &&&12m&C40-3&2016-12-25&44&14.4-488.3&4.7-158.6&33\\
  &&&7m&\nodata&2016-11-05&8&8.9-44.8&2.9- 14.5&46\\
 &&&7m&\nodata&2016-11-06&8&9.5-44.7&3.1- 14.5&46\\
 95.07-98.72&107.07-108.95 &2016.1.00164.S&12m&C40-3&2016-12-23&44&14.5-490.7&4.9-167.0&33\\
\enddata
\end{deluxetable}

\begin{deluxetable}{cccccccc}
\tablecolumns{8}
\tablewidth{0pc}
 \tabletypesize{\scriptsize}
\tablecaption{Imaging Parameters\label{tab:im_param}.}
\tablehead{\colhead{Molecule} &\colhead{Transition}  &\colhead{$f_{\rm rest}$}    &\colhead{beam}  &\colhead{rms} &\colhead{rms}&\colhead{Flux recovery}\\
\colhead{} &\colhead{}  &\colhead{(GHz)}    &\colhead{($\arcsec \times \arcsec$)}  &\colhead{(mJy beam$^{-1}$)} &\colhead{(mK)} &\colhead{(\%)}}
\startdata
  \hline
  $^{13}$CO&$1-0$&110.201&$1.7 \times 1.6$&0.501 &18.6 &$85.5  \pm 1.2$\\
  &$2-1$&220.399&1.7 $\times$1.4    &1.655 &17.5&\\
   C$^{18}$O&$1-0$&109.782&$1.7 \times 1.6$&0.393 &14.7 &$84.4  \pm   1.7$\\
   &$2-1$&219.560&1.7 $\times$  1.4    &0.978 &10.4&\\
   C$^{17}$O &$2-1$&224.714&1.9 $\times$ 1.4    &0.736 &6.7&\\
   CS &$2-1$&97.981&$2.0 \times 1.8$&0.321 & 11.3&$110.8  \pm   2.4$\\
    &$5-4$&244.936&2.1 $\times$ 1.3    &0.842&6.3&\\
     &$6-5$&293.912&1.6 $\times$ 1.3    &1.283&8.4&\\
\enddata
\tablecomments{The flux recovery rate of CCH $N=1-0$ is calculated for the sum of $J=1/2-1/2$ and $J=3/2-1/2$.
Similarly, the flux recovery rate of CH$_3$CN ($5_k-4_k$) is calculated together with that of H41$\alpha$ because they are blended in the single-dish data.
RMS noise is measured for cubes with $\Delta v = 10$ km s$^{-1}$ before the primary beam correction. 
Data of CCH($N=1-0$) and SiO($2-1$) were tapered with the CASA parameter uvtaper = 1arcsec.}
\end{deluxetable}

\begin{deluxetable}{ccc}
\tablecolumns{3}
\tablewidth{0pc}
 \tabletypesize{\scriptsize}
\tablecaption{Coordinates of analyzed positions \label{tab:coords}}
\tablehead{\colhead{Positions} &\colhead{RA (ICRS)}  &\colhead{DEC(ICRS)}  }
\startdata
  \hline
 N1 & 13h37m00.745s&-29d51m43.752s\\
N2 &13h37m00.447s& -29d51m47.855s\\
 N3      & 13h37m00.237s&-29d51m52.979s\\
 N4      & 13h37m00.389s&-29d51m59.933s\\
 N5     &  13h37m00.489s& -29d52m03.400s\\
S1     & 13h37m00.079s&-29d52m16.713s\\
S2 & 13h37m00.418s&-29d52m11.738s\\
S3 & 13h37m00.782s& -29d52m07.462s\\
S4       & 13h37m01.312s& -29d51m59.979s\\
S5      & 13h37m00.892s& -29d51m49.747s\\
C1     &13h37m00.910s&-29d51m55.700s\\
C2 &13h37m00.579s&-29d51m54.873s\\
 \enddata
\end{deluxetable}

\begin{longrotatetable}
\begin{deluxetable}{cccccccccccc}
\tablecolumns{11}
\tablewidth{0pc}
\tablecaption{Column densities at N1-N5 \label{tab:colN}}
\tablehead{\colhead{} & \colhead{N1} & \colhead{} & \colhead{N2} & \colhead{} & \colhead{\
N3} & \colhead{} & \colhead{N4} & \colhead{} & \colhead{N5} & \colhead{}\\
\cline{2-3} \cline{4-5} \cline{6-7} \cline{8-9} \cline{10-11}\\
\colhead{Molecule} & \colhead{N (cm$^{-2}$)} & \colhead{T$_{\rm ex}$ (K)} & \colhead{N (cm$^{-2}$)} & \colhead{T$_{\rm ex}$ (K)} & \colhead{\
N (cm$^{-2}$)} & \colhead{T$_{\rm ex}$ (K)} & \colhead{N (cm$^{-2}$)} & \colhead{T$_{\rm ex}$ (K)} & \colhead{N (cm$^{-2}$)} & \colhead{T$_{\rm ex}$ (K\
)} }
\startdata
$^{13}$CO & $(5.0 \pm 0.4) \times 10^{16}$  & $6.5 \pm 0.3$   & $(5.4 \pm 0.5) \times 10^{16}$  & $8.6 \pm 0.7$   & $(7.1 \pm 0.6) \times 10^{16}$  & $10.2 \pm 0.9$   & $(2.3 \pm 0.3) \times 10^{16}$  & $10.2 \pm 1.4$   & $(3.6 \pm 0.2) \times 10^{16}$  & $8.0 \pm 0.2$  \\
C$^{18}$O & $(1.0 \pm 0.1) \times 10^{16}$  & $6.3 \pm 0.4$   & $(1.0 \pm 0.1) \times 10^{16}$  & $9.5 \pm 1.5$   & $(1.5 \pm 0.1) \times 10^{16}$  & $10.7 \pm 1.4$   & $(4.7 \pm 0.5) \times 10^{15}$  & $12.3 \pm 3.3$   & $(8.0 \pm 0.6) \times 10^{15}$  & $8.5 \pm 0.5$  \\
C$^{17}$O & $(1.4 \pm 0.2) \times 10^{15}$  & ${\it 6.5 \pm ...} $   & $(1.7 \pm 0.1) \times 10^{15}$  & ${\it 9.0 \pm ...}$   & $(2.2 \pm 0.1) \times 10^{15}$  & ${\it 10.0 \pm ...}$   & $(7.6 \pm 0.8) \times 10^{14}$  & ${\it 11.0 \pm ...}$   & $(1.3 \pm 0.1) \times 10^{15}$  & ${\it 8.0 \pm ...}$  \\
CS & $(7.8 \pm 0.2) \times 10^{13}$  & $6.8 \pm 0.1$   & $(8.1 \pm 0.9) \times 10^{13}$  & $8.9 \pm 0.3$   & $(1.0 \pm 0.1) \times 10^{14}$  & $9.1 \pm 0.3$   & $(3.2 \pm 0.3) \times 10^{13}$  & $9.4 \pm 0.3$   & $(3.5 \pm 0.4) \times 10^{13}$  & $7.8 \pm 0.3$  \\
CN & $(1.4 \pm 0.1) \times 10^{14}$  & $4.6 \pm 0.1$   & $(6.4 \pm 0.4) \times 10^{14}$  & $4.1 \pm 0.1$   & $(6.7 \pm 0.3) \times 10^{14}$  & $4.5 \pm 0.1$   & $(3.1 \pm 0.2) \times 10^{14}$  & $4.4 \pm 0.1$   & $(2.7 \pm 0.2) \times 10^{14}$  & $3.9 \pm 0.1$  \\
CCH & $(4.7 \pm 0.4) \times 10^{14}$  & $4.2 \pm 0.1$   & $(7.4 \pm 0.5) \times 10^{14}$  & $5.3 \pm 0.1$   & $(1.0 \pm 0.1) \times 10^{15}$  & $5.6 \pm 0.1$   & $(3.3 \pm 0.3) \times 10^{14}$  & $6.0 \pm 0.2$   & $(2.7 \pm 0.3) \times 10^{14}$  & $4.8 \pm 0.1$  \\
CH$_3$OH & $(7.0 \pm 0.2) \times 10^{14}$  & $8.6 \pm 0.1$   & $(3.9 \pm 0.1) \times 10^{14}$  & $10.2 \pm 0.2$   & $(2.5 \pm 0.1) \times 10^{14}$  & $9.9 \pm 0.2$   & $(5.0 \pm 0.6) \times 10^{13}$  & $9.9 \pm 0.9$   & $(4.9 \pm 0.5) \times 10^{13}$  & $8.7 \pm 0.6$  \\
\enddata
\tablecomments{Excitation temperatures are shown in italic if column densities are calculated using assumed values of excitation temperatures when only one transition was detected.}
\end{deluxetable}
\end{longrotatetable}

\begin{longrotatetable}
\begin{deluxetable}{cccccccccccc}
\tablecolumns{11}
\tablewidth{0pc}
\tablecaption{Column densities at S1-S5 \label{tab:colS}}
\tablehead{\colhead{} & \colhead{S1} & \colhead{} & \colhead{S2} & \colhead{} & \colhead{\
S3} & \colhead{} & \colhead{S4} & \colhead{} & \colhead{S5} & \colhead{}\\
\cline{2-3} \cline{4-5} \cline{6-7} \cline{8-9} \cline{10-11}\\
\colhead{Molecule} & \colhead{N (cm$^{-2}$)} & \colhead{T$_{\rm ex}$ (K)} & \colhead{N (cm$^{-2}$)} & \colhead{T$_{\rm ex}$ (K)} & \colhead{\
N (cm$^{-2}$)} & \colhead{T$_{\rm ex}$ (K)} & \colhead{N (cm$^{-2}$)} & \colhead{T$_{\rm ex}$ (K)} & \colhead{N (cm$^{-2}$)} & \colhead{T$_{\rm ex}$ (K\
)} }
\startdata
$^{13}$CO & $(3.0 \pm 0.6) \times 10^{16}$  & $8.5 \pm 1.5$   & $(2.6 \pm 0.6) \times 10^{16}$  & $6.3 \pm 0.7$   & $(9.7 \pm 1.9) \times 10^{16}$  & $7.1 \pm 0.7$   & $(2.4 \pm 0.3) \times 10^{16}$  & $7.5 \pm 0.7$   & $(3.6 \pm 0.4) \times 10^{16}$  & $6.9 \pm 0.4$  \\
C$^{18}$O & $(6.0 \pm 1.5) \times 10^{15}$  & $9.4 \pm 2.7$   & $(4.2 \pm 1.1) \times 10^{15}$  & $7.0 \pm 1.2$   & $(1.9 \pm 0.4) \times 10^{16}$  & $7.4 \pm 1.2$   & $(3.6 \pm 0.4) \times 10^{15}$  & $8.2 \pm 0.8$   & $(6.5 \pm 0.7) \times 10^{15}$  & $7.2 \pm 0.5$  \\
C$^{17}$O & $(8.9 \pm 0.9) \times 10^{14}$  & ${\it 8.0 \pm ...}$   & $(5.0 \pm 1.7) \times 10^{14}$  & ${\it 6.0 \pm ...}$   & $(2.5 \pm 0.3) \times 10^{15}$  & ${\it 7.0 \pm ...}$   & $(5.3 \pm 0.5) \times 10^{14}$  & ${\it 8.0 \pm ...}$   & $(9.9 \pm 1.7) \times 10^{14}$  & ${\it 7.0 \pm ...}$  \\
CS & $(3.3 \pm 0.6) \times 10^{13}$  & $8.0 \pm 0.4$   & $(3.0 \pm 0.6) \times 10^{13}$  & $6.7 \pm 0.4$   & $(9.3 \pm 0.8) \times 10^{13}$  & $7.1 \pm 0.2$   & $(2.8 \pm 0.5) \times 10^{13}$  & $6.9 \pm 0.3$   & $(2.4 \pm 0.4) \times 10^{13}$  & $6.0 \pm 0.4$  \\
CN & $(9.3 \pm 0.9) \times 10^{13}$  & $4.6 \pm 0.2$   & $(3.3 \pm 0.5) \times 10^{14}$  & $3.3 \pm 0.1$   & $(3.4 \pm 0.2) \times 10^{14}$  & $4.2 \pm 0.1$   & $(4.1 \pm 1.2) \times 10^{14}$  & $3.2 \pm 0.1$   & $(6.0 \pm 1.5) \times 10^{14}$  & $3.1 \pm 0.1$  \\
CCH & $(2.5 \pm 0.4) \times 10^{14.0}$  & $5.2 \pm 0.2$   & $(3.1 \pm 0.8) \times 10^{14.0}$  & $3.8 \pm 0.2$   & $(6.9 \pm 0.6) \times 10^{14.0}$  & $4.5 \pm 0\
.1$   & $(2.7 \pm 0.4) \times 10^{14.0}$  & $4.0 \pm 0.1$   & $(3.5 \pm 1.0) \times 10^{14.0}$  & $3.6 \pm 0.2$  \\
CH$_3$OH & $(2.7 \pm 0.1) \times 10^{14.0}$  & $9.2 \pm 0.2$   & $(3.3 \pm 0.1) \times 10^{14.0}$  & $7.9 \pm 0.1$   & $(4.7 \pm 0.1) \times 10^{14.0}$  & $8.9 \
\pm 0.1$   & $(9.8 \pm 0.6) \times 10^{13.0}$  & $7.8 \pm 0.3$   & $(3.3 \pm 0.5) \times 10^{13.0}$  & $7.7 \pm 0.5$  \\
\enddata
\end{deluxetable}
\end{longrotatetable}

\begin{deluxetable}{cccccc}
\tablecolumns{5}
\tablewidth{0pc}
\tablecaption{Column densities at C1-C2 \label{tab:colC}}
\tablehead{\colhead{} & \colhead{C1} & \colhead{} & \colhead{C2} & \colhead{} \\
\colhead{Molecule} & \colhead{N (cm$^{-2}$)} & \colhead{T$_{\rm ex}$ (K)} & \colhead{N (cm$^{-2}$)} & \colhead{T$_{\rm ex}$ (K)} }
\startdata
$^{13}$CO & $(2.4 \pm 0.2) \times 10^{16}$  & $9.7 \pm 1.1$   & $(6.1 \pm 0.4) \times 10^{16}$  & $9.9 \pm 0.6$  \\
C$^{18}$O & $(4.2 \pm 0.5) \times 10^{15}$  & $11.7 \pm 2.7$   & $(1.5 \pm 0.1) \times 10^{16}$  & $10.9 \pm 0.7$  \\
C$^{17}$O & $(1.1 \pm 0.2) \times 10^{15}$  & ${\it }6.7 \pm ...$   & $(2.5 \pm 0.1) \times 10^{15}$  & ${\it 10.0 \pm ...}$  \\
CS & $(5.6 \pm 0.5) \times 10^{13}$  & $10.0 \pm 0.3$   & $(5.7 \pm 0.8) \times 10^{13}$  & $11.1 \pm 0.6$  \\
CN & $(5.0 \pm 0.2) \times 10^{14}$  & $4.4 \pm 0.0$   & $(9.2 \pm 0.6) \times 10^{14}$  & $4.1 \pm 0.1$  \\
CCH & $(5.7 \pm 0.3) \times 10^{14}$  & $5.3 \pm 0.1$   & $(8.6 \pm 0.5) \times 10^{14}$  & $6.1 \pm 0.1$  \\
CH$_3$OH & $(9.8 \pm 0.8) \times 10^{13}$  & $9.9 \pm 0.4$   & $(2.0 \pm 0.1) \times 10^{14}$  & $10.7 \pm 0.4$  \\
H$_2$CO & $(6.2 \pm 1.3) \times 10^{12}$  & $12.5 \pm 3.5$   & $(9.4 \pm 0.9) \times 10^{12}$  & $16.0 \pm 1.4$  \\
NNH$^+$ & $(3.5 \pm 0.9) \times 10^{12}$  & $6.2 \pm 0.5$   & $(7.7 \pm 0.4) \times 10^{12}$  & $7.1 \pm 0.1$  \\
HCCCN & $<1.3\times 10^{13}$  & $...$   & $(1.7 \pm 0.1) \times 10^{13}$  & $17.3 \pm 1.7$  \\
CH3CCH & $<1.3\times 10^{14}$  & $...$   & $(3.6 \pm 0.2) \times 10^{14}$  & $27.8 \pm 0.6$  \\
\enddata
\end{deluxetable}

\startlongtable
\begin{deluxetable}{cccccccc} 
\tablecolumns{6} 
\tablewidth{0pc} 
\tablecaption{Observed peak intensities and velocity-integrated intensities for detected lines. \label{tab:intens}} 
\tablecomments{Table \ref{tab:intens} is published in its entirety in the machine-readable format.
      A portion is shown here for guidance regarding its form and content. For transitions undetected, upper limit values of 3 $\sigma$ are shown. 
      Peak intensities are taken from a channel smoothed to 10 km s$^{-1}$. We show intensities of data smoothed to the beam size of $2.8''.$}
      The absolute flux calibration has accuracy of about 10 \%. Asterisks are added to values for CS($6-5$) at S1 to indicate that these values
      seem to show more errors than what are theoretically predicted from the primary beam correction, and they need to be taken with caution.
\tablehead{\colhead{Transition} &\colhead{Position}  & \colhead{I$_{\rm peak}$} &\colhead{$\Delta$I$_{\rm peak}$} & \colhead{I $\Delta$V} & \colhead{$\Delta($I $\Delta$V)}\\
\colhead{} &\colhead{} & \colhead{(mK)}& \colhead{(mK)} & \colhead{(K km/s)} & \colhead{(K km/s)} } 
\startdata 
$^{13}$CO($1-0$)               &N1         &   1.2e+03 &   1.2e+01 &   4.8e+01 &   3.7e-01\\ 
   $^{13}$CO($1-0$)               &N2         &   9.0e+02 &   1.2e+01 &   5.4e+01 &   4.2e-01\\ 
   $^{13}$CO($1-0$)               &N3         &   1.7e+03 &   1.2e+01 &   7.1e+01 &   4.2e-01\\ 
   $^{13}$CO($1-0$)               &N4         &   7.6e+02 &   1.2e+01 &   2.3e+01 &   3.4e-01\\ 
   $^{13}$CO($1-0$)               &N5         &   1.3e+03 &   1.2e+01 &   3.3e+01 &   2.9e-01\\ 
   $^{13}$CO($1-0$)               &C1         &   2.7e+02 &   1.2e+01 &   2.5e+01 &   4.9e-01\\ 
   $^{13}$CO($1-0$)               &C2         &   1.7e+03 &   1.2e+01 &   6.0e+01 &   4.3e-01\\ 
   $^{13}$CO($1-0$)               &S1         &   8.6e+02 &   1.3e+01 &   3.8e+01 &   4.2e-01\\
\enddata 
\end{deluxetable} 



\end{document}